\documentclass{emulateapj}

\newcommand{\etal}{et~al.~}
\newcommand{\Msun}{\ifmmode{M_\odot}\else$M_\odot$~\fi}

\newcommand{\kmsMpc}{\hbox{km\thinspace s$^{-1}$\thinspace Mpc$^{-1}$}} 
\newcommand{\TSPH}{{\sc TreeSPH~}}

 \newcommand{\jt}{{\tilde\jmath}}
\newcommand{\be}{\begin{eqnarray}}
\newcommand{\ee}{\end{eqnarray}}

\newcommand{\osix}{O\hspace{0.2mm}VI}

\slugcomment{Received 2002 July 1; accepted 2003 June 1}

\begin{document}
\title{Galaxy Formation: CDM, Feedback and the Hubble Sequence} 
\author{Jesper Sommer-Larsen, Martin G\"otz and Laura Portinari}
\affil{ Theoretical Astrophysics Center,
Juliane Maries Vej 30, DK-2100 Copenhagen {\O}, Denmark}

\begin{abstract}
\noindent
{\TSPH} simulations of galaxy formation in a standard
$\Lambda$CDM cosmology, including star formation, effects of energetic stellar
feedback processes and of a meta-galactic UV field have been performed, 
resulting in a mix of realistic disk, lenticular and elliptical galaxies at 
redshift $z$=0.

The disk galaxies
are deficient in angular momentum by only about a factor of two compared to
observed disk galaxies for simulations with fairly strong
star-bursts in early, proto-galactic clouds, leading to ``blow-away'' of the
remaining gas in the clouds.
In this respect the  present scenario is hence doing 
almost as well as the WDM scenarios discussed by Sommer-Larsen \& Dolgov.
The surface density profiles of the stellar disks are approximately 
exponential and
those of the bulges range from exponential to $r^{1/4}$, as observed.
The bulge-to-disk ratios of the disk galaxies are consistent
with observations and likewise are their integrated $B$--$V$ colours, which 
have
been calculated using stellar population synthesis techniques.  
Furthermore, the observed $I$-band Tully-Fisher relation can be matched, 
provided that the stellar mass-to-light ratio of disk galaxies is
$M/L_I \sim$ 0.8, similar to what was found by Sommer-Larsen \& 
Dolgov from their WDM simulations and in fair agreement with several recent
observational determinations of $M/L_I$ for disk galaxies.

The ellipticals and lenticulars have approximately $r^{1/4}$ stellar surface
density profiles, are dominated by non disk-like kinematics and flattened due 
to non-isotropic stellar velocity distributions, again consistent with 
observations.

Hot halo gas is predicted to cool out and be accreted onto the 
Galactic disk at a rate of 0.5--1 M$_{\odot}$/yr at $z$=0, consistent with 
upper
limits deduced from $FUSE$ observations of \osix. 
We have analyzed in detail the formation history of two disk galaxies with 
circular speeds comparable to that of the Milky Way and find gas accretion 
rates, and hence bolometric X-ray luminosities of the haloes, 6--7 times
larger at $z \sim$ 1 than at $z$=0 for these disk galaxies. More 
generally, it is found that gas infall rates onto these disks are nearly
exponentially declining with time, both for the total disk and the ``solar 
cylinder''.
This theoretical result hence supports the exponentially declining gas infall
approximation often used in chemical evolution models. The infall time-scales 
deduced are $\sim$5--6~Gyr, comparable to what
is adopted in current chemical evolution models to solve the ``G-dwarf
problem''.

The disk of one of the two galaxies forms ``inside-out'', the other
``outside-in'', but in both cases the mean ages of the stars in the outskirts
of the disks are $\ga$6--8 Gyr, fairly consistent with the findings of
Ferguson \& Johnson for the disk of M31.

The amount of hot gas in disk galaxy haloes is consistent with observational 
upper limits. The globular cluster M53 and the LMC are ``inserted'' in the 
haloes of the two Milky Way like disk galaxies and dispersion measures to 
these objects calculated. The results are consistent with upper limits 
from observed dispersion measures to pulsars in these systems. 

Finally, the results of the simulations indicate that the observed peak in the
cosmic star formation rate at redshift $z \sim 2$ can be reproduced.
Depending on the star formation and feedback scenarios
one predicts either a cosmic star formation rate which decreases monotonically
with redshift beyond
these redshifts or a second peak at $z$$\sim$6--8 corresponding to the
putative population III, and interestingly similar to recent estimates of the
redshift at which the Universe was reionized. These various scenarios
should hence be observationally constrainable with upcoming instruments
like JWST and ALMA.    
\end{abstract}
\vspace{1cm}
\vspace{6cm}

\newpage

\section{Introduction}
\label{s:intro}

The formation of galactic disks is one of the most important unsolved
problems in astrophysics today. In the currently favored hierarchical
clustering framework, disks form in the potential wells of dark matter
haloes as the baryonic material cools and collapses dissipatively.  It
has been shown (Fall \& Efstathiou 1980) that
disks formed in this way can be expected to possess the observed
amount of angular momentum (and therefore the observed spatial extent), but 
only under the condition that the infalling gas retains most of its original 
angular momentum --- see also Fall (2002).

Numerical simulations of this collapse scenario in CDM cosmologies, however,
have so far consistently indicated that, when only radiative cooling processes
are
included, the infalling gas loses too much angular momentum (by more than 
an order of magnitude) and the resulting disks are accordingly much
smaller than required by the observations (e.g., Navarro \& Benz 1991;
Navarro \& White 1994; Navarro, Frenk, \& White 1995; Navarro \& Steinmetz 1997
using N-body/SPH codes and Bryan 2003 using an Adaptive Mesh Refinement 
code, Bryan 1999).
This discrepancy is known as the {\em angular momentum problem} of
disk galaxy formation.
It arises from the combination of the following two facts:
a) In the CDM scenario the magnitude of linear density fluctuations  
$\sigma(M) = <(\delta M/M)^2>^{1/2}$ increases steadily with decreasing
mass scale $M$ leading to the formation of non-linear, virialized structures
at increasingly early epochs with decreasing mass, i.e.~the hierarchical
``bottom-up'' scenario.  b) Gas cooling
is very efficient at early times due to gas densities being generally
larger at higher redshifts, as well as the rate of inverse Compton cooling 
also 
increasing very rapidly with redshift ($\propto (1+z)^4$).  a) and b)
together lead to rapid condensation of small, dense gas clouds,
which subsequently lose energy and (orbital) angular momentum by dynamical
friction against the surrounding dark matter halo before they
eventually merge to form the central disk.
A mechanism is therefore needed that prevents, or at least delays,
the collapse of small proto-galactic gas clouds and allows the gas to
preserve a larger fraction of its angular momentum as it settles into
the disk. Weil, Eke, \& Efstathiou (1998) 
have shown that if the early radiative
cooling is suppressed (by whatever means), numerical simulations can
indeed yield more realistically sized disks --- see also Eke, Efstathiou \&
Wright (2000).
The physical mechanism by which cooling is suppressed or
counteracted, however, was not specified.

Sommer-Larsen \etal (1999, hereafter SLGV99) discussed the effects of 
various
stellar reheating mechanisms in more detail using numerical
\TSPH simulations of disk galaxy formation in the SCDM cosmology.
They found that reheating of the Universe resulting
from more or less {\it uniformly} distributed star formation
does not lead to a solution to the angular momentum (AM) problem, but that 
localized
star-bursts in proto-galactic gas clouds might: {\it if} the star-bursts can
blow the remaining bulk part of the gas out of the small
and dense dark matter haloes associated with the clouds, then test simulations
showed that the gas later settles gradually
forming an extended, high angular momentum disk galaxy in the
central parts of a large, common dark matter halo. 
The physics of global gas blow-out processes were considered in early 
calculations by
Dekel \& Silk (1986) and Yoshii \& Arimoto (1987), indicating
that star-bursts might well blow out most of the gas in small galaxies with
characteristic circular speed (defined in this paper as the circular speed in
the disk at 2.2 exponential scalelengths) $V_c~\la$~100 km/s. More recent, 
detailed simulations by Mac Low \& Ferrara
(1999) suggest that this global blow-out scenario may not work so
well in {\it disk} galaxies, even in small ones: the star-bursts typically 
lead to bipolar
outflows of very hot gas perpendicular to the disk of the small galaxy,
only expelling a minor fraction of the disk gas. This, however, is due to
the particular geometry of a disk galaxy (the thinness and flatness of the 
disk);
in a more roundish and bulky, proto-galactic dwarf galaxy, one would expect
the energetic effects of star-bursts on the bulk of the dwarf galaxy gas
to be much stronger.

It has proven difficult to realistically implement thermal energy feedback in 
cosmological N-body/SPH codes incorporating radiative cooling (Thacker \& 
Couchman 2000 and references therein; Navarro \& Steinmetz 2000;
SLGV99). As first discussed by Katz (1992) this is mainly due to problems
in resolving a multi-phase inter stellar medium (ISM) at the scales of 
individual star burst regions
in cosmological simulations using SPH. It is also related to the smoothing of
discontinuities in the thermodynamic variables inherent in the SPH method
--- some improvement on this, at least in relation to shocks, may be gained by
replacing thermal energy by entropy as an independent variable in SPH 
(Springel \& Hernquist 2002). In recent years, however, some progress
has been made on the implementation of thermal energy feedback in SPH 
simulations (Gerritsen 1997; Mori \etal 1997; Thacker \& Couchman 2000, 2001;
Springel \& Hernquist 2003 --- see also Yepes \etal 1997; Ferrara \& Tolstoy 
2000; Lia, Portinari \& Carraro 2002; Semelin \& Combes 2002;
Governato \etal 2002; Abadi \etal 2003 and Meza \etal 2003).

As was shown by Sommer-Larsen \& Dolgov (2001, SLD01) by going from the
CDM structure formation scenario to the warm dark matter (WDM) one can
alleviate (and possibly even completely overcome) the AM
problem without invoking effects of energetic feedback in the simulations at
all (this does not imply, of course, that feedback necessarily {\it is}
unimportant; it should rather be seen as a minimal assumptions approach). 
Fine-tuning of the warm dark matter particle mass to about 1 keV is
required, however. In contrast, the salient feature about ``conventional'' CDM 
is that as long as the dark matter particles are much heavier than one keV, 
the actual particle mass does not matter for structure formation (note though,
that axions, despite being ultra-light, behave like CDM).   

In this paper we show how a mix of realistic disk, lenticular and elliptical 
galaxies\footnote{Pictures of some of the galaxies can be seen at 
http://www.tac.dk/\~~\hspace{-1.4mm}jslarsen/Hubble\hspace{0.1mm}\_ \hspace{-0.7mm}Sequence}
can be obtained in fully cosmological ($\Lambda$CDM), gravity/hydro 
simulations invoking star formation, energetic stellar feedback processes in
early, proto-galactic clouds and a meta-galactic UV field. This is achieved
by treating (albeit in a coarse way) the gas in regions where star-bursts are 
returning energy to the
ISM as a two-phase medium, consisting of a ``cold'' ($T \sim 10^4$~K) and a hot
($T \sim 10^6$~K) component. Related, recent work includes Thacker \& Couchman
(2001), Steinmetz \& Navarro (2002) and Springel \& Hernquist (2003).
    
In section~\ref{s:feedback} the implementation of stellar feedback processes
in the simulations is discussed. Section~\ref{s:methods} gives a short 
presentation of the numerical code 
and the initial conditions. The simulations themselves are briefly
described in section~\ref{s:simulations}, and the results obtained are 
analyzed and
discussed in section~\ref{s:results}. Finally, in section~\ref{s:conclusions}
we summarize our conclusions and present a brief outlook.

\section{Star formation and stellar feedback}
\label{s:feedback}
The star formation efficiency (SFE, defined as the ratio between the dynamical
timescale and the star formation timescale --- see below) in the Galactic disk
is quite small at 
present, at most a few percent (e.g., Silk 1997). After extensive 
experimentation we have found
that such a low SFE in combination with a ``conventional'' initial mass 
function
(IMF) produces too low an energy feedback rate in the early and dense 
proto-galactic 
clouds to drive the remaining gas out of the (cold) dark matter potential wells
in which they have formed. As a consequence, such a low
SFE at early times does not lead to a solution of the AM problem for the
CDM + feedback scenario - see section~\ref{s:results}. A similar result was
found by Abadi \etal (2003), but we note that Governato \etal (2002)
found for one CDM galaxy formation simulation, that the AM problem could be 
solved (to within about a factor of two) using low, early SFE + feedback. 
In any case, for that scenario one is still faced
with the problem that the stellar haloes become too massive (Governato \etal
2002) and/or the stellar spheroids too dominant and centrally 
concentrated (Abadi \etal 2003; Meza \etal 2003). Moreover,
Governato \etal find for their CDM simulation that the number of stellar 
satellite galaxies is too large compared to observations of the Milky Way 
and M31 (the ``missing satellites problem'').

To solve the AM problem (to within about a factor of two; 
section~\ref{s:results}) as well as the other problems mentioned above 
(section~\ref{s:results}; Sommer-Larsen \etal 2003)
we find that considerably larger energy feedback 
rates are required at early times. This can be achieved, for example, by
invoking at early times 1) a considerably larger SFE or/and 2) a
substantially more top-heavy IMF. In this paper, we shall not 
attempt to give a detailed, theoretical substantiation of these
two possibilities, but a few, brief plausibility arguments can be given as 
follows: 

The
stars formed in the early, proto-galactic clouds will have low heavy element
abundance, as they will belong to the first generations of stars formed in the
Universe. Hence a physical motivation for case 1) could be related
to the fact that the strength of stellar winds decreases with decreasing 
metallicity (Kudritzki \etal 1989; Kudritzki \& Puls 2000), 
possibly with a sharp drop 
below metallicities of about 1/100 solar (Kudritzki 2002).
So for {\it increasing} metal abundance, the stronger stellar winds may 
regulate star formation so as to {\it decrease} the SFE, due to the energy
and momentum feedback to the star forming, molecular clouds. It is also 
possible 
that the SFE could depend on the thermal history of the gas such that gas which
has been heated to temperatures well above $T_{crit}$$\sim$$10^4$~K 
(below which
atomic radiative cooling becomes unimportant) and subsequently cooled down
again has a lower SFE than gas which has never been heated above $T_{crit}$.
Physically this could be related to a possible multi-phase structure of the
former type of gas, effects of magnetic fields, destruction of molecules etc.

With respect
to case 2) some theoretical work on star formation indicates that the
IMF of the first generations of stars was in fact more top-heavy than present
day IMFs (Abel, Bryan \& Norman 2002; Bromm, Coppi \& Larson 2002;
and references in Chiosi 2000).
Padoan \& Nordlund (2002) find that the high-mass slope of the IMF
becomes steeper at early times (when the magnetic fields in molecular clouds
were
considerably weaker than the $B \sim \mu$G at present), but also that the 
low-mass cut-off of 
the IMF at the same time increases considerably such as to effectively make 
their proposed IMF more top-heavy (or more appropriately ``bottom-light'') at 
early times ({\AA}.~Nordlund, 2002, private communication).

In this paper we consider case 1), but we have no reason to believe
that case 2) would not work equally well.
 
We use two distinctly different star formation and feedback modes depending
on the thermal history of the star forming gas (as chemical evolution is not
yet implemented in the simulations):

\medskip \noindent
$\bullet$ ``Early'' star formation mode

\medskip \noindent
Cool, dense gas, which has {\it always} been cooler than $T_{crit}
\simeq 10^4$ K is assumed to form stars rapidly, i.e. on a timescale 
comparable to the local, dynamical time $t_{dyn}$. We express this star 
formation timescale as
\begin{equation}
t_{*,e} = \frac{t_{dyn}}{\epsilon_e} = \frac{1}{(4 \pi G \rho_{gas})^{1/2}}
\frac{1}{\epsilon_e} ~~,
\end{equation}
where the star formation efficiency $\epsilon_e \sim 1$ (we find that the 
specific value of $\epsilon_e$ used is not critical for the outcome of the 
simulations as long as $\epsilon_e \sim 1$; we have used $\epsilon_e=1$ in
this work).
Such fast star formation is assumed to be triggered when the gas density of an
SPH particle
exceeds a certain critical value, chosen to be $n_{\rm{H,e}}=0.3$ cm$^{-3}$;
we have experimented with other values and found that the outcome of the
simulations is very robust to changes in this parameter. In a star formation
event an SPH particle is converted fully into a collisionless star particle
of the same mass, thereby conserving the total number of particles in the
simulation. The star particle remains a star for the rest of the simulation ---
non-instantaneous recycling is being implemented in a forthcoming version of 
the code. The moment of conversion from SPH to star
particle is, as is customary, determined by a probabilistic approach. 

The triggering of an SPH particle for star formation may (A) or may not (B)
trigger a burst of self-propagating star formation (SPSF) in the cold, dense 
gas
surrounding it: in scenario ``A'' not only the SPH particle which gets above
the critical density threshold, but also its neighbouring cold and dense
SPH particles with densities above $n_{\rm{H,e,low}} (< n_{\rm{H,e}})$
are triggered for conversion into star particles on their individual, 
dynamical timescales. Such self-propagating star formation is observed at 
present in some star-burst galaxies (e.g., in expanding super-shells --- 
see Mori \etal 1997). SPSF is likely to have been more common at early
times (when metallicities were lower and the ISM more homogeneous) --- see
McCray \& Kafatos (1987). We stress that the above implementation
of SPSF is just a simple, schematic way of modelling complicated 
(sub-resolution) physical processes.

For scenario ``A'' we ran 4 series of simulations with $n_{\rm{H,e,low}}$= 
0.05,
0.1, 0.2 and 0.25 cm$^{-3}$ respectively, corresponding to the conversion of
approximately 5, 4, 3 and 2\% of the gas mass initially in the simulation into
stars at $z \ga 5-6$; we dub this the ``population III'' scenario. 
The 4 different values of $n_{\rm{H,e,low}}$ were used to control the strength
of the star-bursts to check whether the outcome of the simulations was
sensitive to the choice of this parameter. In particular we wanted to check
whether considerable fine-tuning is required to solve the AM problem. Early
star-bursts converting more than about 5\% of the initial gas into stars
lead to halo star metallicities which are too large compared to observations
(see section 5.2 and also SLGV99), so this sets an upper limit to the strength
of the early star-bursts. 

In scenario ``B'' only the initial SPH particle above the critical density 
threshold
is triggered for star formation on the dynamical time scale --- we dub this the
``population II'' scenario.

When a star particle is formed, it is assumed to represent a population of 
stars born at 
the same time in accordance with a Salpeter IMF. It will hence feed energy
back to the local ISM: during the first 5 Myr
only stellar winds are assumed to contribute, during the subsequent about
35 Myr also (and more importantly) type II supernova explosions as well 
(the lightest stars which explode as SNII have a 
mass of about 8 $M_{\odot}$ and a lifetime of about 40 Myr). 
Only stars with masses greater than 30 $M_{\odot}$ are assumed to contribute
significant amounts of energy to the ISM through winds, and such stars are
assumed to deposit a total of 10$^{50}$ ergs per star over their lifetime.
Stars with masses greater than 8 $M_{\odot}$ are assumed to deposit 10$^{51}$ 
ergs per star to the ISM as they explode as SNII at the end of their lifetime.
The energy from the ``star-burst'' is deposited in the ISM as thermal energy 
at a constant rate over the lifetime of the burst, which is 
taken to be the above mentioned about 40 Myr (Mori \etal 1997). 
The burst energy is fed back to the, at any given time ($<$40 Myr), 50 nearest 
SPH particles using the smoothing kernel of Monaghan \& Lattanzio (1985). 
Part of the thermal energy is subsequently converted into kinetic energy (by
the code) as the resulting ``super-shell'' expands. While a star
particle is feeding energy back to its neighbouring SPH particles, radiative 
cooling
of these is switched off - this is an effective way of modelling with SPH a 
two-phase 
ISM consisting of a hot component ($T \sim 10^6-10^7$ K) and a much cooler
component ($T \sim 10^4$ K) --- see Mori \etal (1997); Gerritsen (1997)
and Thacker \& Couchman (2000, 2001).
Scenario ``A'' above 
typically leads to complete ``blow-away'' of the remaining gas in the
proto-galactic ``dwarf'' galaxy hosting the star-burst, whereas scenario ``B''
is much more gentle, but still leads to a considerable lowering of the density
of the remaining gas.

\medskip \noindent
$\bullet$ ``Late'' star formation mode

\medskip \noindent
The gas which forms the disks of disk galaxies through smooth, dilute and 
fairly well ordered cooling flows has typically been heated to temperatures
well above $T_{crit}$ and has subsequently cooled down and settled onto the 
disk to become potentially star forming. Such gas is assumed to form stars on
a much slower timescale  
\begin{equation}
t_{*,l} = \frac{t_{dyn}}{\epsilon_l} ~~,
\end{equation}
where $\epsilon_l \ll $1. 
As discussed previously, such a large change of SFE between early and late
phases of galaxy formation is in our experience 
required to form disks with the densities, radial and vertical structures, 
cold gas fractions, sizes and morphologies of observed disk galaxies at 
present. A value of $\epsilon_l$=0.0025 was adopted in order
to obtain (cold) gas fractions in the disk galaxies at $z=0$ consistent with 
observational estimates (e.g., Sommer-Larsen 1996: the gas fraction
(by mass) in Sa-Sc galaxies is $\sim$0.10-0.25 --- the mean of our 28 SPSF 
disk galaxy runs (see section 4) is 0.17$\pm$0.01\%, the dispersion 0.07). 
This adopted value of $\epsilon_l$
is somewhat low; we expect it to increase when non-instantaneous recycling of
gas is properly taken into account in the simulations. Star formation through
this ``channel'' is assumed to take place down to a quite low density 
threshold of 0.01 cm$^{-3}$. This, however, is still large
enough to ensure that the gas has cooled down to a temperature 
$T$$\sim$10$^4$ K,
at which the radiative cooling function is effectively truncated. For the
(typically disk forming) star particles formed through this slow mode feedback
was completely switched off. This is a crude, but effective way of modelling 
in our simulations 
the results of Mac Low \& Ferrara (1999) obtained at much higher 
resolution, that for star-bursts in a well-organized, cold gas {\it disk} only
a small fraction of the feedback energy is deposited in surrounding cold gas.
 
Finally we note that eq.~(2) implies a star formation law in the continuous
limit (and above the density threshold) of the form 
$\dot{\rho}_{*} \propto \rho_{\rm{cg}}^{1.5}$, where
$\rho_{\rm{cg}}$ is the cold ($T$$\sim$10$^4$ K) gas density.
The exponent of 1.5 is both physically and empirically (at $z$$\simeq$0) well 
motivated (Kennicutt 1998).
\section{The code and the initial conditions}
\label{s:methods}

\subsection{The code}
\label{s:code}

We use the gridless Lagrangian $N$-body and Smoothed Particle
Hydrodynamics code {\TSPH} described in SLGV99.
Our {\TSPH} code is modelled after that of Hernquist \& Katz (1989).
The spline smoothing kernel of Monaghan \& Lattanzio (1985) is used
throughout for smoothing of gas dynamical properties as well as softening
of gravity, as in Hernquist \& Katz (1989). Individual time-stepping,
implemented in a way similar to that of Springel, Yoshida \& White 
(2001), is used in the code. The system time-step is always the
smallest of all individual time-steps.

We include radiative gas cooling and heating in the simulations.
The radiative heating corresponds to a redshift dependent, homogeneous and 
isotropic UV background radiation field produced by AGNs and young galaxies.
This meta-galactic UV field is modelled after Haardt \& 
Madau (1996) --- the UV field switches on at redshift $z \sim$ 6.
The radiative cooling function is that of a primordial gas, modified by the
effects of the UV field as discussed by, e.g., Vedel \etal (1994). 
The code furthermore incorporates inverse Compton cooling, which is also
explicitly redshift-dependent. Star formation is incorporated by converting
SPH particles into star particles by the schemes described in the
previous section. The sum of the numbers of SPH and star particles is kept
constant.

The smoothing length of each SPH particle is
adjusted so as to keep the number of neighbors close to~50.

Finally, we have in this work used the shear-free Balsara (1995) 
viscosity, rather than the standard Monaghan-Gingold (1983) viscosity 
used previously.

\subsection{The initial conditions}
\label{s:ic}

Our cosmological initial conditions are based on a $\Lambda$CDM model with
($\Omega_M,\Omega_{\Lambda}$)=(0.3,0.7) and a Hubble parameter
$H_0 = 100h\,\kmsMpc = 65\,\kmsMpc$, resulting in a present age of the 
Universe of 14.5 Gyr. Following Eke, Cole, \& Frenk (1996)
we normalize the spectrum to $\sigma_8(z=0)=1.0$,
where as customary $\sigma_8^2$ is the mass variance within spheres of
comoving radius $8h^{-1}$~Mpc, extrapolated from the linear regime of
perturbation growth. 

First, a cosmological $N$-body simulation (dark matter only)
with 128$^3$ particles, a comoving box length of 10~$h^{-1}$Mpc and
starting at a redshift of $z_i$=39 was run using the Hydra code
(Couchman \etal 1995). At z=0, gravitationally bound dark matter
haloes were identified by running a friends-of-friends algorithm with 
linking length 0.2. 

Second, 12 dark matter haloes (see section~\ref{s:simulations})
selected for detailed galaxy formation simulations were resampled and
resimulated: For each dark matter halo a galaxy forming sub-volumne was 
determined by tracing all dark matter particles, which at $z$=0 were inside 
of 4~$r_{200}$ ($r_{200}$ being the radius inside of which the mean dark 
matter density is 200 times the critical), back to the initial conditions
at $z_i$=39. The initial, co-moving linear extent of the galaxy forming 
sub-volume was $\sim$1.5-3~$h^{-1}$Mpc.
At $z_i$, SPH particles were added to the galaxy 
forming sub-volume, one SPH per dark matter (DM) particle, with a mass of $f_b$
times the original DM particle mass, where $f_b$ is the baryonic 
fraction (see below). The masses of the DM particles in the sub-volume
were reduced accordingly to ($1-f_b$) times the original mass.
In the surrounding region of the cosmological simulation (with co-moving linear
extent of about 10~$h^{-1}$Mpc) no SPH particles were
added and the dark matter particles were increasingly coarsely resampled with 
increasing distance from the galaxy forming sub-volume. This, by now customary,
procedure has been described by many authors, e.g., Navarro \& White 
(1994); Gelato \& Sommer-Larsen (1999) and Thacker \& Couchman (2000).
The galaxies were selected to be at least 1 Mpc away from
galaxy groups and 0.5 Mpc away from larger galaxies at $z$=0. They were also
selected such as not to undergo major merging events (with companion to
parent mass ratios of 1:3 or more) since $z$=1.

We use a baryonic mass fraction $f_b=0.10$, consistent with
nucleosynthesis constraints ($0.015 h^{-2} \la \Omega_b \la 0.02
h^{-2}$) and with the
observationally determined baryonic fractions in galaxy groups and clusters
(e.g., Ettori \& Fabian 1999).
The masses of the SPH (and star) particles and the high
resolution DM particles in the galaxy forming sub-volume are $4.0\times 10^6$ 
and $3.6\times 10^7~h^{-1}\Msun$, respectively.
The SPH particles are assigned an initial thermal energy corresponding
to a temperature $T_i\simeq 100$~K.
Gravitational interactions between particles are softened
according to the prescription of Hernquist \& Katz (1989), with
softening lengths of $\tilde{\epsilon}_{g,*}$=1.3 $h^{-1}$kpc for the gas 
and star particles and $\tilde{\epsilon}_{DM}$=
2.8 $h^{-1}$kpc for the high resolution DM particles. The gravitational 
softening lengths were kept fixed in co-moving coordinates until $z$=2.9, 
then constant in physical coordinates.

To check for effects of numerical resolution, one series of disk galaxy 
simulations was run with half these softening lengths, and one additional
simulation was
run with one quarter of these softening lengths and eight times higher mass
resolution. For these latter simulations the softening lengths were kept 
fixed in co-moving coordinates until $z$=6.7, then constant in physical 
coordinates.

\section{The simulations}
\label{s:simulations}

We selected 12 dark matter haloes from the cosmological N-body simulation
for the detailed galaxy formation runs. The masses of these haloes span
about a factor of 10, with 4000--45000 dark matter particles inside 
$r_{200}$ at $z$=0 and circular velocities at $r_{200}$ ranging
from about 100 to 185 km/s. After resampling the galaxy formation 
simulations
consisted of 30000-150000 SPH+DM particles. We started out by running all
12 galaxy simulations using the self-propagating star formation prescription
(mode ``A'' in section~\ref{s:feedback}) with a lower density threshold of
$n_{H,e,low}$=0.1 cm$^{-3}$ (and $n_{H,e}$=0.3 cm$^{-3}$, as always).
Seven of the resulting galaxies had at $z$=0 distinctly disk galaxy like
morphologies and kinematics, the remaining 5 were lenticular (S0) or elliptical
like. Four additional series of simulations were
subsequently run for the 7 disk galaxies: three using again early
SPSF with $n_{H,e,low}$=0.05, 0.2 and 0.25 
cm$^{-3}$ and one series with fast, early, but non self-propagating star 
formation (mode ``B'' in section~\ref{s:feedback}). Typically, $\sim$30000
system time-steps were required to run from $z_i$=39 to $z$=0.
To test for possible numerical effects we ran one additional (and more
time consuming) series of 7 disk galaxy simulations with $n_{H,e,low}$=0.2 
cm$^{-3}$, but with 
gravitational softening lengths of 0.65~$h^{-1}$kpc and 1.4 $h^{-1}$kpc 
for the SPH (and star) and dark matter particles, respectively --- the outcome
of these
simulations compared fairly well with the results of the same simulations with
``standard'' gravitational softening lengths. We shall briefly discuss this in
section 5.9. About 50000 system time-steps were required to run to $z$=0 for
these simulations.

In addition we re-simulated one of the disk galaxies at an 8 times
higher mass resolution (a very CPU intensive run). This
was achieved by extracting the same region from cosmological initial
conditions with identical parameters, but on a $256^3$ mesh rather
than the $128^3$ grid used so far. Identical low-wavenumber Fourier
modes were used for both grids, up to the Nyquist wavenumber of the
coarser mesh. Above that, additional high-wavenumber modes were added
to the finer mesh (up to its Nyquist wavenumber) to account for the extra
small scale power in this simulation.

This procedure resulted in SPH (and star) and high resolution DM
particle masses of $4.9\times 10^5$ and $4.4\times 10^6$ $h^{-1} M_\odot$,
respectively.
Moreover, we used a four times higher force resolution for this simulation
corresponding to gravitational softening lengths of 0.33 and 0.69~$h^{-1}$kpc 
for the SPH
(and star) and high resolution DM particles, respectively. The simulation was
an SPSF run with $n_{H,e,low}$=0.25 cm$^{-3}$ and the result is compared to
that of the ``standard'' resolution simulation of the same galaxy (also with  
$n_{H,e,low}$=0.25 cm$^{-3}$) in section 5.9. This run required $\sim$80000
system time-steps.

Finally, to illustrate the importance of having a large initial SFE we ran
the 7 ``normal resolution'' disk galaxy simulations with a constant, low
SFE of $\epsilon$=0.02. They were of the non SPSF type and run with the 
standard gravitational softening lengths. 
 
In total we ran 12+4$\times$7=40 ``standard'' galaxy formation simulations,
7+1 additional at higher resolution and 7 with a constant, low SFE. The 
computational costs amounted to about 5 years worth of single processor CPU 
time on a SGI Origin 2000 computer.

\section{Results}
\label{s:results}

All results presented correspond to the final state of the simulations at
redshift $z=0$ unless otherwise explicitly mentioned. A general presentation
of the results of the simulations at higher redshift will be given in a 
forthcoming paper. 

\subsection{Surface density profiles, specific angular momenta and structural 
parameters}

As mentioned in the previous section, 7 of the 12 galaxies simulated have disk
galaxy like morphologies and kinematics, with the bulk of the stars on 
approximately circular orbits in a disk, most of the rest of the stars 
in an inner, bulge-like component and finally a small fraction in a round and 
dynamically insignificant stellar halo surrounding the galaxies. The disk
galaxies formed in our simulations are hence qualitatively quite similar to
observed disk galaxies like the Milky Way. Of the remaining 5 galaxies, two
have a minor fraction of the stars on nearly circular, disk orbits; we classify
these as lenticulars (S0s) and the remaining three galaxies have no disks
at all; we classify these as ellipticals --- see further below.

The disk galaxies have approximately exponential stellar disk surface density
profiles and exponential to $r^{1/4}$ bulge profiles, all in good agreement
with observations. Two examples of a disk galaxy stellar surface density
profile are shown in Figure~1 (for all stars within 2 kpc vertical distance 
from the disk) - the surface density of cold gas ($T \sim 10^4$ K) is also
shown. The stellar profiles of the lenticular and elliptical galaxies are not 
exponential, as shown in Figure~2, but approximately follow an $r^{1/4}$ law, 
as shown in Figure~3.  

\begin{figure}
\plotone{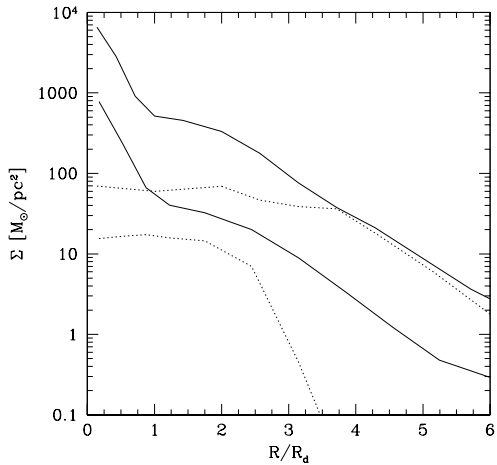}
\caption{Surface density profiles for two Milky Way like disk galaxies, denoted
S1 and S2 in this paper. Radial coordinate is in units of the exponential disk 
scalelength at $z$=0. Top curves correspond to S1; stars are marked by solid 
line, cold gas ($T$$\sim$$10^4$~K) by dotted. We estimate the (stellar)
bulge-to-disk ratio of this galaxy to B/D=0.17$\pm$0.05. Bottom curves 
correspond to S2 --- the curves have been displaced 1 dex downward for clarity
(line-types as for S1). S2 has B/D=0.27$\pm$0.08.
\label{f:1}}
\end{figure}

\begin{figure}
\plotone{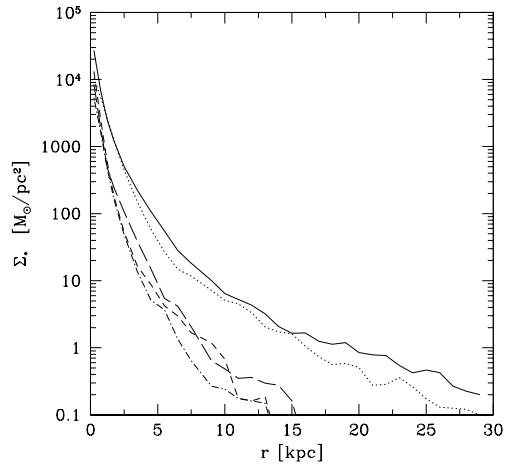}
\caption{Stellar surface density profiles of the three elliptical 
and two lenticular galaxies.
\label{f:2}}
\end{figure}

\begin{figure}
\plotone{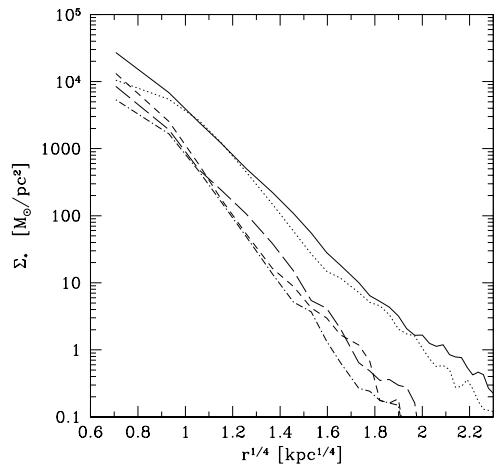}
\caption{Same as in Fig.~\ref{f:2}, but versus $r^{1/4}$. 
\label{f:3}}
\end{figure}

The bulges of the disk galaxies are generally confined to being within a 
radius $r_B \sim 1-1.5$ kpc from the centers. Stellar disk scale lengths
were determined by fitting exponential surface density profiles to the 
region of the disk outside of the bulge, $R > r_B$.

Bulge-to-disk ratios were determined by extrapolating the exponential disk 
profiles, obtained as described above,
to the center of the galaxies. Using these decompositions (which
make no assumptions about the bulge surface density profiles) the specific 
angular momenta of the stellar {\it disks} were estimated taking explicitly
into account also the region with overlap between disk and bulge. 
As $r_B$ is comparable to the gravitational softening lengths one has to 
check how affected the structural decomposition parameters are by gravity
softening effects - we return to this at the end of the subsection. 

Characteristic circular speeds $V_c$ for the disk galaxies were calculated
using the approach of SLD01, but as an addition taking into account also the 
dynamical effect of the bulges. 

In Figure~4 we show the ``normalized'' specific angular momenta 
$\jt_* = j_*/V_c^2$ of the final disks formed in all
35 disk galaxy simulations as a function of $V_c$. 
As argued by SLGV99 one expects $\jt_*$ to 
be almost independent of $V_c$ on both theoretical
and observational grounds. Also shown in the figure is the median 
``observed'' value of $\jt_*$, calculated as in SLGV99 and SLD01
for a Hubble parameter of 65 km/s/Mpc, together with the observational 
1-$\sigma$ and 2-$\sigma$ limits. As can be seen from the figure, the
specific angular momenta of the stellar disks lie only about a factor of three
below the observed median. This is about an order of magnitude better than
what is obtained in similar CDM simulations not invoking 
stellar feedback processes, as discussed by many authors (e.g.,
Navarro, Frenk, \& White 1995; SLGV99).

\begin{figure}
\plotone{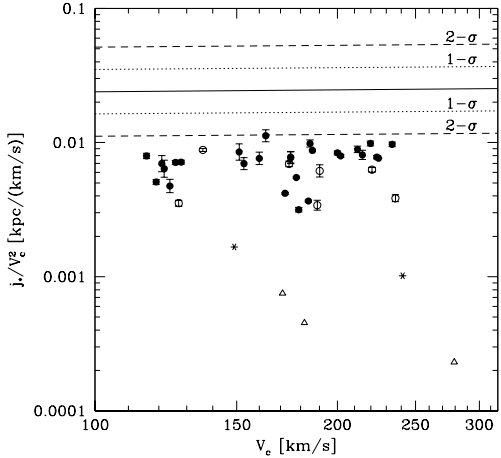}
\caption{Normalized stellar specific angular momenta at $z$=0, 
$\jt_* = j_*/V_c^2$, of the galaxies formed in the simulations:
Disk galaxies formed in the 4 series of runs with early, self-propagating star
formation (SPSF) are shown by filled circles. The lenticulars (S0) and 
ellipticals (formed in the SPSF simulations with 
$n_{{\rm H,e,low}}$=0.10 cm$^{-3}$) are shown
by star symbols and open triangles, respectively. Disk galaxies formed in 
simulations without self-propagating star formation are shown by open circles.
The solid line shows the median value from the
observational data on disk galaxies of Byun (1992), obtained as 
explained in SLD01 (but in this paper assuming $H_0$=65 km/s/Mpc),
the dotted and dashed lines bracket the 1-$\sigma$
and 2-$\sigma$ intervals around the median.
\label{f:4}}
\end{figure}

Also shown in the figure are the normalized specific angular momenta of the
two lenticular and three elliptical galaxies. The ``effective'' $V_c$ for
these has been calculated using the stellar mass versus $V_c$ relation
for the disk galaxies (Figure~14). The specific angular momenta of the E/S0s
are about an order of magnitude smaller than those of the disk galaxies, 
broadly consistent with observations (e.g., Navarro \& Steinmetz 1997).

For the remainder of this paper we shall be concerned with the properties of 
the disk galaxies. We will analyze the E/S0s in detail in a forthcoming paper.

In Figure~5 we show the normalized specific angular momenta of the 28 disk 
galaxies from the SPSF simulations versus the
dimensionless spin-parameter $\lambda_{200}$ of their dark matter haloes
($\lambda \equiv J|E|^{1/2}/GM^{5/2}$ is evaluated at $r_{200}$).
Fall \& Efstathiou (1980) proposed that the angular momentum of a disk
galaxy should be proportional to the spin-parameter of its dark matter halo,
and indeed a positive correlation is seen in Figure~5, although the scatter
is considerable (the linear correlation coefficient is 0.51, so 
the probability that
$\jt_*$ and $\lambda_{200}$ are uncorrelated is less than 3\%). 
Filled triangles, squares, pentagons and circles correspond to 
$n_{H,e,low}$ = 0.05, 0.10, 0.20 and 0.25 cm$^{-3}$, respectively. It is clear
from the figure that there is no statistically significant evidence that
one particular choice of the parameter $n_{H,e,low}$ leads to a better 
solution of the AM problem than the other. Hence, for these SPSF simulations
extreme fine-tuning of the amount of energetic feedback seems not to be 
required.    

\begin{figure}
\plotone{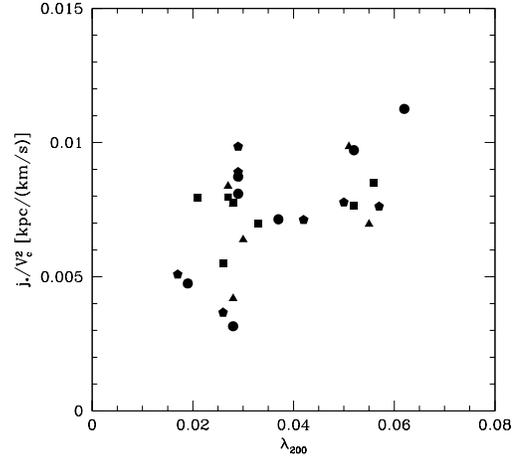}
\caption{Normalized disk stellar specific angular momenta versus the
spin parameter $\lambda_{200}$ (evaluated at $r_{200}$ of the dark matter
halo) for the simulations with self-propagating star formation (SPSF). Filled
triangles, squares, pentagons and circles correspond to 
$n_{H,e,low}$ = 0.05, 0.10, 0.20 and 0.25 cm$^{-3}$, respectively.
\label{f:5}}
\end{figure}

\begin{figure}
\plotone{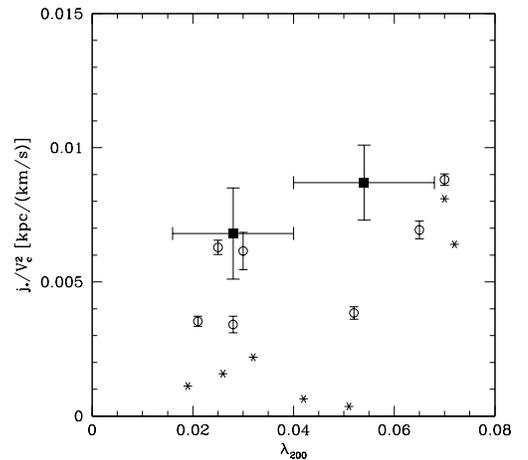}
\caption{As Fig.~\ref{f:5}, but for the disk galaxies 
formed in the 7 simulations without SPSF (shown by the open circles). Also
shown (by filled squares) are the data for the 28 disk galaxies from 
Fig.~\ref{f:5} binned into two bins. The vertical bars on these last two data 
points are the dispersions in the two bins; the horizontal show the bin size.
Finally, shown by star symbols are the results of the 7 constant, low SFE
($\epsilon$=0.02) disk galaxy runs - no B/D decomposition was attempted for
these galaxies, because they in general are very small and concentrated.  
\label{f:6}}
\end{figure}

In Figure~6 we show the normalized specific angular momenta of the 7 disk
galaxies from the simulations without SPSF versus $\lambda_{200}$. Also shown
are the results from Fig. 5 binned into two bins. There is clearly an 
indication that the simulations without SPSF are not doing as well 
in terms of solving AM problem. We also show the results of the 7 constant,
low SFE ($\epsilon$=0.02) disk galaxy runs - these are in general giving
very poor results in terms of solving the AM problem: the mean, spin-parameter
corrected (see below), normalized angular momentum of these 7 runs is just 
27$\pm$6\% of that of the 28 SPSF runs. 

The spin-parameters of our 7 disk galaxy haloes are on average somewhat smaller
than the median found in cosmological N-body simulations --- see below. This 
is most likely just an effect of small number statistics. 
Assuming (simplisticly) a linear relationship between $\jt_*$ and 
$\lambda_{200}$ on the basis of the trend seen in Fig. 5 we correct the 
normalized specific angular momenta of the 28 disk galaxies formed in the SPSF 
simulations as follows to see how well we can expect to do in solving the AM
problem. We assume   
a theoretical median value of $\lambda$ of 0.05 (Barnes \&
Efstathiou 1987; Heavens \& Peacock 1988) and hence multiply
the normalized specific angular momenta of 28 disk galaxies by 
(0.05/$\lambda_{200}$). The results are shown in Figure~7. 
It is seen that the disk galaxies formed in SPSF simulations are really
only deficient in specific angular momentum by about a factor of two relative 
to the observed median.
The feedback+CDM scenario presented in this paper is hence doing almost as
well as the WDM scenario discussed by SLD01 in terms of solving the AM problem
for disk galaxies. 

\begin{figure}
\plotone{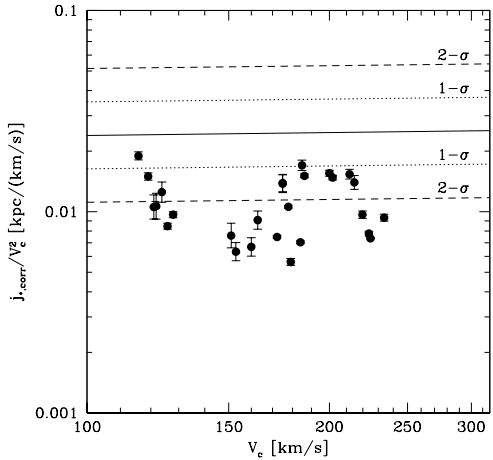}
\caption{Spin-parameter corrected normalized specific angular
momenta for the 28 disk galaxies formed in the simulations with SPSF. 
Line symbols as in Fig.~\ref{f:4}.
\label{f:7}}
\end{figure}

Although we are making an appropriate comparison between ``observational''
{\it disk} specific angular momenta and model {\it disk} specific angular 
momenta (both obtained through B/D decomposition) we note that some authors 
present total (disk+bulge) specific angular momenta for simulated galaxies.
For comparison with
these works we calculated $j_{*,tot}$ for the 35 model disk galaxies and
found $<j_{*,tot}> = 0.87 <j_*>$, so the difference between the two types
of specific angular momenta is quite small. 

In Figure~8 we show the exponential scale-lengths of the 35 disk galaxies
versus $V_c$. Not surprisingly, these are also too short by about a factor
of two relative to the observed median (taken from SLGV99, but corrected to 
$h$=0.65).
In Figure~9 we show the same, but
corrected for spin-parameter effects, as discussed above, and only for the
SPSF simulations, given the result shown in Figure~6. 

\begin{figure}
\plotone{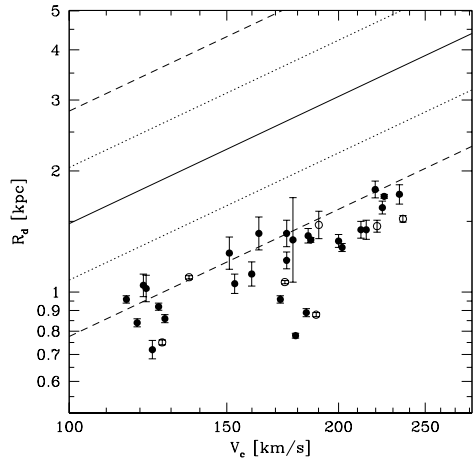}
\caption{Exponential scale-lengths of all 35 disk galaxies (symbols as in 
Fig.~\ref{f:4}). Line symbols as in Fig.~\ref{f:4}.
\label{f:8}}
\end{figure}

\begin{figure}
\plotone{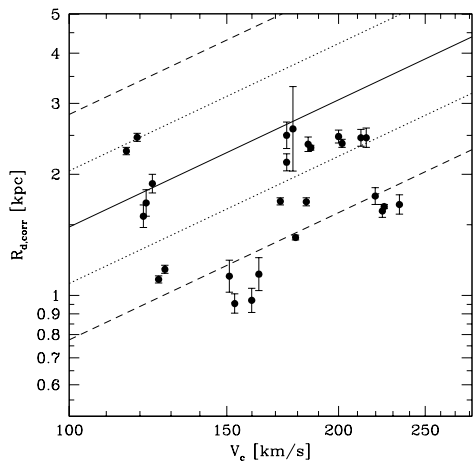}
\caption{Spin-parameter corrected exponential scale-lengths of
the 28 disk galaxies formed in the runs with early, self-propagating star
formation --- see text for details. Line symbols as in Fig.~\ref{f:4}.
\label{f:9}}
\end{figure}

In order to compare our model disk galaxies to observations some indicator of
Hubble type $T$ is required (as usual $T$=1 corresponds to Sa, $T$=2 to Sab,
$T$=3 to Sb and so on). A seemingly obvious choice is the bulge-to-disk ratio 
$B/D$ given the well-known correlation between the $B$--band bulge-to-disk
ratio $(B/D)_B$ and galaxy type. What we have available from the decomposition
of the model galaxies are {\it mass} $B/D$. Byun (1992) 2-D decomposed
$I$-band images of $\sim$~1000 Sb-Sd galaxies (3$ \le T \le $7) from the sample
of Mathewson \etal (1992). Contrary to $(B/D)_B$, $I$-band $B/D$ trace
the mass $B/D$ better (e.g., Byun 1992). In Figure~10 we show the
mean $(B/D)_I$ for the $\sim$~1000 disk galaxies in Byun's sample.
The bars correspond to the dispersion in $(B/D)_I$ for each Hubble type.
Clearly, it is virtually impossible to classify a disk galaxy on the
basis of its $I$-band (and hence approximately {\it mass}) $B/D$ (at least for
$3 \le T \le 7$). 

\begin{figure}
\plotone{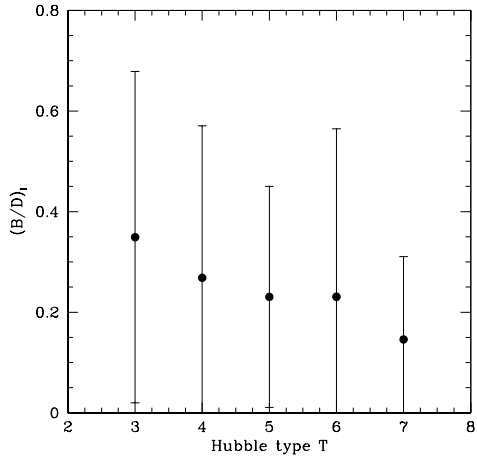}
\caption{Mean $I$-band bulge-to-disk ratios versus Hubble Type $T$
for the $\sim$~1000 Sb-Sd galaxies for which Byun (1992) performed 2-D
$B/D$ decomposition. The bars correspond to the dispersion in each bin.
\label{f:10}}
\end{figure}

Another potential disk galaxy type diagnostic is the 
so-called birthrate parameter, $b$, defined as the ratio of the present to 
past average star formation rate, 
{\mbox{$b=SFR/< SFR >$}}, 
for the {\it disk} stars --- see Kennicutt, 
Tamblyn \& Congdon (1994) and references therein. In Figure~11 we show
the mean $b$ versus Hubble type ($T$=1--7) for the data given in Kennicutt 
\etal (1994) after 3-$\sigma$ ``clipping''.  
The bars correspond to the dispersion in $b$ for each Hubble type. Clearly,
$b$ is a much better diagnostic of Hubble type than the mass $B/D$. A linear
fit to the data yields 
\begin{equation}
b = 0.19 \, T - 0.13  ~~ (1 < T \la 5)~~,
\end{equation}
which is the relation we will use in the following when comparing our model 
disk galaxies to observations.

\begin{figure}
\plotone{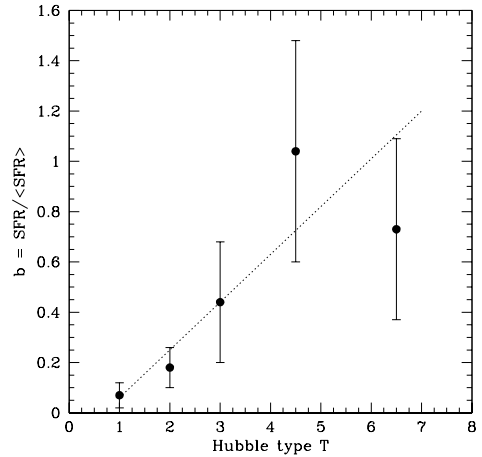}
\caption{Mean birthrate parameter $b$ versus Hubble Type $T$. The original data
have been taken from Kennicutt \etal (1994) and the data shown have 
been 3-$\sigma$ ``clipped''. The bars correspond to the dispersion in each bin
and the dotted line is the linear fit to the data used in this paper (eq. 3).
\label{f:11}}
\end{figure}

Figure~12 shows the stellar {\it mass} bulge-to-disk ratios $B/D$ of the 35 
disk galaxies versus the birthrate parameter $b$. As can be seen from Fig.~11
$b \sim 0.1$ for Sa's increasing to $b \sim 1$ for Sc's. 
The lack of trend seen in Fig.~12 is consistent with the weak trend found 
observationally for $(B/D)_I$ (and also for $K$-band $B/D$ ratios, e.g., de 
Jong 1996), when allowing for the large dispersion in each
bin seen in Fig.~10. 

\begin{figure}
\plotone{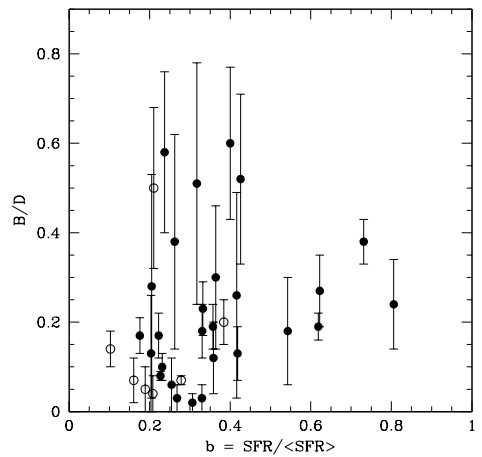}
\caption{Bulge-to-disk ratios of all 35 disk galaxies versus
birthrate parameters --- see text for details (symbols as in Fig.~\ref{f:4}).
\label{f:12}}
\end{figure}

As mentioned previously, since $r_B$ is comparable to the gravitational 
softening lengths, the structural bulge/disk decompositions can be
affected by spurious effects of gravity softening (e.g., Bate \& Burkert 1997;
Sommer-Larsen, Vedel \& Hellsten 1998a,b). To test for
such effects we performed the following experiments: Switching star formation
off and slowly reducing the softening lengths to final values half of the
initial (which were 1.3 $h^{-1}$kpc for the gas (and star) particles and 
2.8 $h^{-1}$kpc for DM particles), we continued the 7 disk galaxy SPSF runs 
with $n_{H,e,low}$ = 0.20 
cm$^{-3}$ for one additional Gyr, starting at $z$=0. As the orbital time 
scale in the bulge and the main disk is considerably less than
1 Gyr the angular momentum distribution of the collisionless components
(stars and dark matter) will be approximately preserved, as angular
momenta are adiabatic invariants. 

We then performed the bulge/disk decomposition of the resulting 7 disk 
galaxies as before and found that the mean $B/D$ increased by 0.20$\pm$0.07
(up from 0.20 to 0.40), the mean $R_d$ by --0.07$\pm$0.16 kpc and the mean 
disk $j_*$ by 38$\pm$73 kpc km/s (up from 229 to 267 kpc km/s)
as the softening lengths were reduced by a factor of two. We do not expect
any significant further changes if the softening lengths were reduced even
more since now $\tilde{\epsilon}_{g,*} < r_B$ and the central gravitational 
potential
is completely dominated by stars and gas. So a statistically significant 
increase in the mean $B/D$ ratio results from going to smaller softening 
lengths, bringing it in better agreement with the recent estimate of $\la$0.6 
for the {\it mass} $B/D$ ratio independent of disk galaxy type 
(Portinari, Sommer-Larsen \& Tantalo 2003). As a consequence
the resulting {\it disk} $j_*$ would be expected to increase somewhat. This
is indeed hinted by the above results though the statistical significance
of this can not be established on the basis of the present 7 experiments.
The disk scale lengths appear to be only slightly overestimated (by 
6$\pm$14\%).  

\subsection{Broad band $B$--$V$ colours of the disk galaxies}

Besides the birthrate parameter, a variety of other characteristics of 
observed galaxies is known to depend
on their Hubble type (see e.g.\ the review by Roberts \& Haynes 1994).
In this Section we discuss the integrated $B$--$V$ colours of our simulated 
disk galaxies.

We estimate their colours ``post-process'':
First we calculate the average chemical enrichment history 
of each galaxy on the basis of its star formation history (SFH),
inferred from the age distribution of the stars in the final galaxy 
(inside of $r$=30 kpc) at $z=0$.
Once the metallicity evolution $Z(t)$ is computed, we derive from the star 
formation rate $\Psi(t)$ of the galaxy its
integrated luminosity $L$ in any given passband as:
\begin{equation}
\label{integL}
L = \int_0^{t_G} \Psi(t) \, L_{SSP}(t_G-t,Z(t)) \, dt
\end{equation}
where $t_G$ is the age of the galaxy and $L_{SSP}(\tau,Z)$ is the luminosity 
of a Single Stellar Population (SSP; Tinsley 1980; Renzini \& Buzzoni 
1986) of age $\tau$ and metallicity $Z$. 
In the following we give further details on the individual ingredients in
calculating the $B$--$V$ colours.
\subsubsection{The chemical evolution history}

The metal enrichment is calculated post-process, with the purpose of
obtaining an approximate estimate of the integrated $B$--$V$ colours of the
disk galaxies. 

Some level of global pre-enrichment 
is expected from the initial ``fast'' star formation activity responsible 
for the early feedback effects (see Section~2).
This early star formation is localized in the cool gas clouds 
but otherwise widely
distributed over the simulation volume;
it typically peaks at redshifts $z$$\sim$6--8 (for the SPSF runs --- see
Section 5.4), followed 
by a more gentle level of activity. 
A second phase of star formation, associated with the formation of the main 
galaxy itself, commences at redshifts $z \sim 4-5$. As a first approximation
one can hence divide the star formation and chemical enrichment history into 
two phases: an early phase resulting in fairly uniform metal (pre)enrichment 
of the simulation volume and a second corresponding to the formation of the
main galaxy.

The level of pre-enrichment is calculated with a chemical
evolution model representing a closed-box with the same SFH as that of the
entire simulation volume. We use the chemical
evolution model by Portinari, Chiosi \& Bressan (1998), suitably modified 
so as to treat the SFH $\Psi(t)$ as an input information, rather than 
calculating its own star formation rate after a prescribed Schmidt-like 
law. A Salpeter 
Initial Mass Function (IMF) is assumed throughout the calculations, 
with logarithmic slope 1.35 and mass limits 0.1---100~\Msun.
The metallicity of the closed-box at $t=$1.5~Gyr ($z=4.2$) is taken as the 
level of pre-enrichment for the subsequent second phase of main galaxy
formation.
We obtain levels of pre-enrichment in the range 1/100 to 1/20~$Z_{\odot}$
(increasing as the strength of the early star-bursts increases, causing a 
larger fraction of gas to be converted into stars during this early phase
--- see Section 2), with 1/40~$Z_{\odot}$ as the typical value.

Starting from the appropriate initial metallicity level, we then calculate
the chemical evolution of the main galaxy based on its SFH during the remaining
13~Gyr (from 1.5 to 14.5~Gyr). The SFH is directly provided by 
the age distribution of the star particles that reside in the galaxy
at the end of the simulation.
For the chemical evolution of the main galaxy, the closed-box assumption 
would be
too crude an approximation: not all the baryons that end up in the final 
galaxy are immediately and equally available for star formation since 
the beginning, as the closed-box model would presume. Realistically, 
what happens is that gas progressively cools out and becomes available 
for star formation ($T \sim 10^4$ K, $n_H \ga 0.01$ cm$^{-3}$). We can regard 
this ``cool out'' as a sort of ``infall history'' for the galaxy.
We estimate the accretion history of cold gas onto the main
galaxy as follows. At each time $t$, we know the total amount
of cold gas in the simulation volume, $M_{cg}^{tot}(t)$, as well as the total
star formation rate $\Psi^{tot}(t)$, and the SFH of the stars that
end up in the galaxy, $\Psi(t)$. The rate at which cold, dense gas becomes 
available
for the star formation relevant to the main galaxy, is simply
approximated by:
\begin{equation}
M_{cg}(t) = M_{cg}^{tot}(t) \, \frac{\Psi(t)}{\Psi^{tot}(t)} 
\end{equation}
Hence we calculate the chemical evolution of the galaxy assuming an infall
model with a mass accretion history given by:
\begin{equation}
M(t) = M_{cg}(t) + M_*(t) 
\end{equation}
where $M_*(t)$ is the mass of the stars in the final galaxy 
that were formed up to time $t$. To do so, we further adapted the chemical 
evolution model by 
Portinari et~al.\ (1998) so that it could treat not only the SFH, 
but also the accretion history 
as input data, in place of the usual prescription
\mbox{$\dot{M} \propto e^{-\frac{t}{\tau}}$}.

Post-process calculation of chemical evolution introduces some
inconsistency with respect to the results of dynamical simulations, since
the chemical model accounts for the gas re-ejection from dying stars, 
which is neglected in the simulations. Hence, for the same SFH 
the two models will end up with a somewhat different gas versus star content. 
For the Salpeter IMF adopted here, a SSP restitutes $\sim$30\% of its mass
to the gaseous phase over a Hubble time, hence this is the order of magnitude
of the discrepancy between the final mass in stars in the chemical model
and in the dynamical simulation.
A consistent calculation of the metallicity distribution of the stars 
in the galaxy would require chemical evolution to be
implemented in the simulations, so that each star particle is consistently 
labelled with its own age and metallicity and can be individually treated 
as a SSP for the sake of colour calculation. This is currently being
implemented in the code. The above mentioned mismatch in final mass in stars, 
however, mostly
affects the integrated magnitude of a galaxy, while its colour is only
marginally sensitive to the slightly different metal enrichment for different 
gas fractions. Hence, for the purpose
of this paper our post-process estimate of the colours is adequate.
%
\subsubsection{The calculation of ``integrated'' $B$--$V$ colours}

The chemical evolution calculated above provides us with the metal enrichment
history $Z(t)$ of the galaxy, which combined with its known SFH $\Psi(t)$
allows us to calculate its integrated luminosity 
from 
Eq.~(\ref{integL}). As to $L_{SSP}$, we adopt the magnitudes and colours
of the SSPs by Tantalo et~al.\ (1996).

The $B$--$V$ colours, calculated as described above, do not include the effects of 
dust and gas re-processing of the stellar radiation, which can be crucial
especially for the youngest stellar populations. In particular,
there is evidence from studies of combined dust extinction and IR 
re-emission that the youngest stellar populations can be heavily obscured 
by the parent clouds for periods of up to
a few $10^7$yr (e.g.\ Silva et~al.\ 1998; Charlot \& Fall 2000). 
In the current paper we deal with this effect in the following simple way: We 
calculate the $B$--$V$ colour for each galaxy
by including all stars formed to the present (corresponding to no
``dust correction'' at all) and by dropping the light contribution of the 
stars formed in the last $2 \times 10^7$~yr; we 
consider the range of colours we obtain as our uncertainty and denote the mean
of the two values ($B$--$V$)$_0$.
As to the extinction from the diffuse gas phase, the reddening effect
on $B$--$V$ for spiral disks is mild (e.g.\ Boissier \& Prantzos 1999) and
moreover the observed colours from Roberts \& Haynes which we use for 
comparison
are the RC3 catalogue ``intrinsic'' ($B$--$V$)$_0$ colours, which have been 
corrected for internal extinction.

Further uncertainties and systematic effects on the colours 
may come from the choice of the IMF, which influences the level
of chemical enrichment. Dynamical arguments favour 
a rather low M/L ratio in spiral
galaxies, both for external galaxies and for our Milky Way (Sommer-Larsen 
\& Dolgov 2001 and references therein --- see also section 5.3), implying a 
more bottom-light IMF
than the standard Salpeter one adopted here. A variety of studies in fact
suggest the the slope of the IMF gets much shallower than the Salpeter value
below 1~\Msun (e.g.\ Kroupa \etal 1993; Chabrier 2003).
In the present paper, however, we consistently use a Salpeter IMF 
since our main purpose is to discuss the {\it trend} of the $B$--$V$ colours
with respect to other galactic properties, such as Hubble type etc.

Finally, systematic differences in the broad-band colours of SSPs
among different authors exist, while better agreement is found for the 
differential variation of SSP colors with age and  metallicity
(e.g.\ Charlot, Worthey \& Bressan 1996; Kodama \& Arimoto 1998). 
So, given the above-mentioned sources of uncertainty
(dust, IMF and theoretical SSPs) the {\it trend} of broad-band 
colours with the properties 
of simulated galaxies is to be considered more robust than their absolute 
values. 

The $B$--$V$ colours we obtain are compared to the observed ones from
the review by Roberts \& Haynes (1994), which is based on a large sample of
galaxies compiled from the RC3 catalogue. We use the disk galaxy birthrate 
parameter $b$ as our morphology indicator and convert from the observational
morphological type $T$ to $b$ as described at the end of Section~5.1 (using 
eq.~3).

In Figure~13 we show the predicted ($B$--$V$)$_0$ colour for the 35 disk 
galaxies
versus the $b$ parameter. Also shown are the observational results from 
Roberts \& Haynes. The agreement between theory and observations
is certainly satisfactory.

\begin{figure}
\plotone{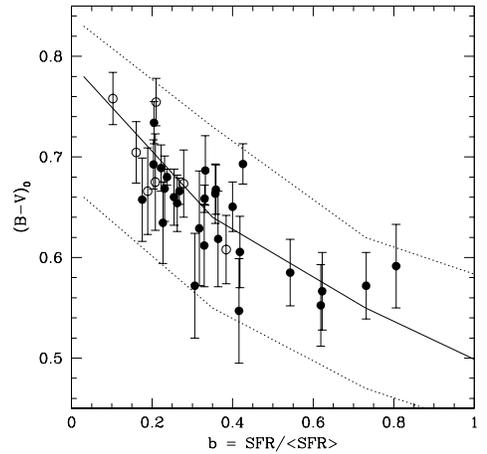}
\caption{Integrated ($B$--$V$)$_0$ colours of all 35 disk galaxies, calculated
using stellar population synthesis techniques --- symbols as in Fig.~\ref{f:4}.
Also shown (solid line) are the mean observational values for disk galaxies 
from Roberts \& Haynes (1994) and the observational 1-$\sigma$ limits 
(dotted line). 
\label{f:13}}
\end{figure}

\subsection{The Tully-Fisher relation and mass-to-light ratios of disk 
galaxies}

In Figure~14 we show the stellar mass of the final disk galaxies formed in 
35 runs versus the characteristic circular speed. Also shown is the $I$-band 
Tully-Fisher
relation (TF) of Giovanelli \etal (1997a) for $h$=0.65, converted
to mass assuming (stellar) mass-to-light ratios $M/L_I$ = 0.5, 1.0 and 2.0
in solar units (used throughout) and applying an 0.2 mag. offset 
(Giovanelli \etal 1997b) to take into account that the typical disk 
galaxy in our 
sample is of type Sab. The
slope of the ``theoretical'' TF matches that of the observed one very well
for a constant mass-to-light ratio, which is required to be $(M/L_I) 
\sim$ 0.8. This fairly small value required is consistent
with the findings of SLD01 for their WDM simulations. Such a low value is
fairly consistent with recent dynamically and/or lensing estimated 
mass-to-light ratios for disk galaxies
(e.g., Vallejo, Braine \& Baudry 2002; Trott \& Webster 2002),
the mass-to-light of the Milky Way (SLD01) and can be obtained from stellar
population synthesis models provided an IMF somewhat more top-heavy (or
more appropriately: more bottom-light) than the Salpeter law is used
(Portinari, Sommer-Larsen \& Tantalo 2003). The
observed
chemical enrichment of galaxy clusters and the global cosmic star formation 
and enrichment history seem to require an IMF with similar properties (Pagel
2002).

\begin{figure}
\epsscale{1.1}
\plotone{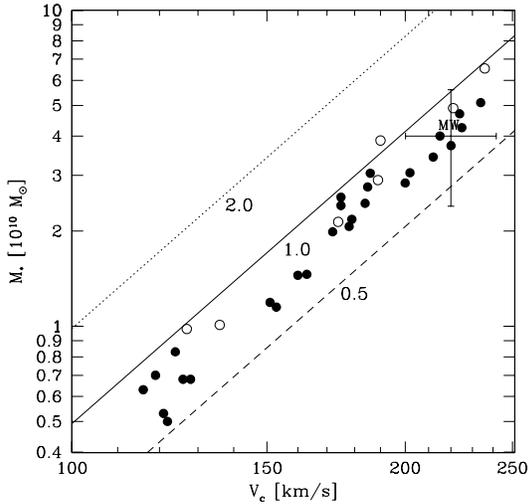}
\caption{The stellar mass vs. circular velocity ``Tully-Fisher'' relation
for the galaxies from the 35 simulations resulting in disk galaxies (symbols 
as in Fig.~\ref{f:4}).
Also shown is the observed $I$-band TF relation for Sc galaxies of
Giovanelli \etal (1997) converted to mass assuming ($M/L_I$)=0.5 
({\it dashed line}), 1.0 ({\it solid line}) and 2.0 ({\it dotted line}) 
and applying an 0.2 mag. offset to correspond to Sab galaxies (Giovanelli 
\etal 1997b), which is the mean type of our simulated disk galaxies.
Finally, the symbol ``MW'' with errorbars shows the likely range of the 
stellar mass and characteristic circular velocity of the Milky Way (from SLD01,
but corrected to stellar mass, rather than total baryonic mass).
\label{f:14}}
\end{figure}

\subsection{Disk galaxy gas accretion rates}
We have calculated the rates at which hot halo gas cools out and subsequently
is deposited onto the disks of the 35 disk galaxies in our sample.
This is of considerable interest because observational upper limits can be
placed on the accretion rate of gas onto the Galactic disk and potentially
also other disk galaxies.

As hot halo gas (at $T_{{\rm max}} \sim 10^6$ K) starts to cool out
near the disk and finally is deposited onto it, at some much lower
temperature ($T_{{\rm min}} \sim 10^4$ K) highly, but not fully, ionized atoms 
like \osix~will be
present in the transition region. Denoting the accretion column density of
H atoms per unit time by $\dot{N}_H$ and the initial (hot halo gas) H density
$n_{H0}$ and neglecting heat conduction, one can show that the column
density of an ion $Z^i$ produced between temperatures $T_{{\rm min}}$ and 
$T_{{\rm max}}$
is
\begin{equation}
N^i_Z = k \frac{\dot{N}_H}{n_{H0}} A_Z \int_{T_{{\rm min}}}^{T_{{\rm max}}} 
(\frac{3}{2} +
s) \frac{\chi}{\chi_e} \frac{f_i dT}{(n_H/n_{H0}) \Lambda} ~~,
\end{equation} 
where $k$ is the Boltzmann constant, $A_Z$ is the abundance of the element
relative to hydrogen, $s$ is 1 for isobaric and 0 for isochoric cooling,
$\chi$ is the number of particles per H atom, $\chi_e$ the number of electrons
per H atom, $f_i$ is the fraction of atoms of element $Z$ which are in
ionization stage $Z^i$, $n_H$ is the H density of the cooling, transition-phase
gas, and $\Lambda$ is the cooling function, such that $\Lambda n_H n_e$,
where $n_e$ is the free electron number density, is the energy loss rate per 
volume (Edgar \& Chevalier 1986, EC). For isobaric cooling (which, cf.
EC, is the most relevant case for the present problem) EC find 
{\mbox{$N$(\osix)=3.8$\cdot$10$^{14}$ cm$^{-2}$}} for 
$\dot{N}_H/n_{H0}$=10$^7$ cm/s 
and assuming solar gas abundance and composition.

The result does not change much for hot halo gas 
oxygen abundances as low as 1/10 solar: 
Galactic halo stars are observed to be $\alpha$-element enhanced, reflecting
that the heavy elements in the stars were primarily produced by type II 
supernovae. For halo stars [O/H]=--1.0 corresponds to [Fe/H]$\simeq$--1.4
(e.g., Pagel 1997). One would expect that the hot halo gas is 
$\alpha$-element
enhanced as well, since the heavy elements in the gas most likely originate
from a) the early phase of halo star formation and/or b) star-bursts in the
Galactic disk. In the latter case the heavy elements produced by the type II
supernovae are transported to the halo via fountain flows (e.g., Mac Low
\& Ferrara 1999). For [Fe/H]$\simeq$--1.4 the radiative (collisional
ionization equilibrium) cooling function is almost an order of magnitude less
than for [Fe/H]=0.0 (e.g., Sutherland \& Dopita 1993). Hence it follows
from eq.~(7) that $N$(\osix) should be similar for 1/10 solar and solar gas 
oxygen abundance (for the same $\dot{N}_H/n_{H0}$). Even if the gas has solar
iron-element abundance {\it and} is $\alpha$-element enhanced with 
[O/Fe]$\simeq$0.4 due to SNII driven fountain outflows from the disk the 
estimate of $N$(\osix) (at a given $\dot{N}_H/n_{H0}$) should not increase by
more than about 50\% because of the increased (radiative) cooling efficiency 
resulting from the $\alpha$-element enhancement. 

From \osix~ absorption lines in the spectra of $\sim$ 100 AGNs 
observed with 
{\it FUSE (Far Ultraviolet Spectroscopic Explorer}), Blair Savage and 
collaborators find 
$N$(OVI)~sin$|b|$~$\simeq$~1.8$\cdot$10$^{14}$~cm$^{-2}$ and
$\simeq$~1.3$\cdot$10$^{14}$~cm$^{-2}$ in the Northern and Southern Galactic 
hemispheres, respectively (Savage \etal 2003). 
Assuming {\mbox{$n_{H0} \sim$~10$^{-3}$ cm$^{-3}$}}
(EC; B.~Savage, 2002, priv. comm.; also what is found from our 
simulations) and combining both sides of the disk, this yields a gas accretion
rate of 2.9$\cdot$10$^{-3}$ 
M$_{\odot}$/yr/kpc$^2$ locally (``locally'' for the observed, \osix~absorbing 
gas actually corresponds to a fairly large region of the disk out to
about 5 kpc from the sun, B.~Savage, 2002, priv. comm.).
Assuming a characteristic size of the Milky Way's gas disk of $R$=15-20 kpc and
that the above local accretion rate is typical, one obtains an estimated total
accretion rate of about 2.0-3.6 M$_{\odot}$/yr 
with an uncertainty of at least a factor of two.
Part (and perhaps most) of this is not likely to stem from hot halo gas 
cool-out, but from fountain flows (B. Savage, 2002, priv. comm.).

In Figure~15 we show the present day disk gas 
accretion rates of the 35 disk galaxies versus their characteristic circular
speed. It is seen that disk galaxies with characteristic circular speed
comparable to that of the Milky Way ($V_c \simeq$ 220 km/s) are found to have
accretion rates of about 0.5--1 M$_{\odot}$/yr at $z$=0, broadly consistent 
with the above observational upper limit. 

\begin{figure}
\plotone{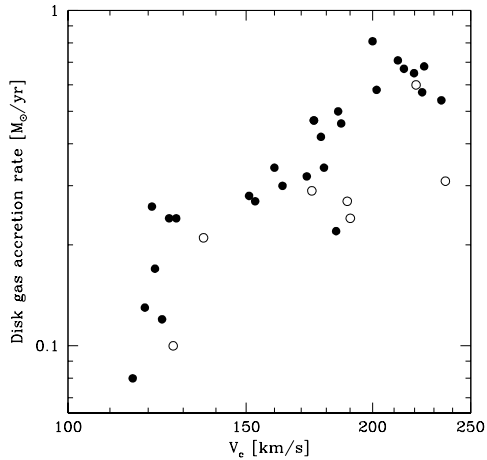}
\caption{Disk gas accretion rates at $z$=0 for the 35 model disk 
galaxies (symbols as in Fig.~\ref{f:4}).
\label{f:15}}
\end{figure}

As an alternative source of gas infall, in particular in relation to the
``G-dwarf problem'' (see section 5.6), the HI high-velocity clouds (HVC)
have been considered. We note that the accretion rate predicted by our models
is comparable
to the maximum Tosi (1988) could obtain with HVCs under a number of
favorable assumptions. With the ``new'' source of gas infall discussed in this
paper (cooled-out halo gas) it is no longer required to appeal to HVC's as
the primary source of gas infall onto the Galactic disk. 

More generally, 
one can show that the rate at which hot gas cools out in the halo is 
proportional 
to $L_{{\rm X,bol}}$$<\frac{1}{T}>$, where $L_{{\rm X,bol}}$ is the
bolometric X-ray luminosity and $<\frac{1}{T}>$ is the 
mean, emissivity 
weighted inverse temperature of the hot halo gas (Toft \etal 2002).
Only few relevant X-ray observations of external disk galaxies are currently
available, but forthcoming XMM-Newton and Chandra observations of nearby,
nearly edge-on, isolated and undisturbed disk galaxies should greatly 
improve the situation and enable the determination of gas accretion rates for
external disk galaxies from the bolometric X-ray luminosity of their hot
haloes (Toft \etal 2002).  

For the simulations presented in this paper we
have assumed that the gas has primordial abundance. If the gas abundance is
1/3 solar, as in the intra-cluster medium, it follows from the work of
Toft \etal that the bolometric X-ray luminosities and hence gas accretion 
rates will be a factor of 3-4 {\it lower} at $z$=0, so for a Milky Way like
disk galaxy about 0.1--0.3~M$_{\odot}$/yr would be expected, even more
consistent with the upper limit for the Galactic disk, deduced from \osix~ 
observations.

\subsection{Disk formation: ``inside-out'' versus ``outside-in''}
It is frequently advocated that in hierarchical structure formation scenarios,
such as the CDM scenario, the formation of galactic disks proceeds 
``inside-out''
in the sense that an initial formation of the inner parts of the
disk is followed by a phase of gradual increase of the size of the disk 
lasting for a Hubble time (e.g., White \& Frenk 1991;
Sommer-Larsen 1991; Mo, Mao \& White 1998).
With the models presented here we are in a position to
test this ``inside-out'' paradigm. In two of the 12 dark matter haloes used in
this work disk galaxies with circular speeds comparable to that of the Milky
Way are formed. Of the total of eight
SPSF simulations run with these two haloes, we analyze for the larger 
halo the one with $n_{H,e,low}$=0.25 cm$^{-3}$ and for the smaller halo the 
one with $n_{H,e,low}$=0.20 cm$^{-3}$ --- the other SPSF simulations of 
these two haloes give similar results, but the runs selected are the ones
which result in the largest $j_*$, being 19 and 11\% larger than the mean 
for the four SPSF runs per halo, respectively. The disk galaxy formed
in the largest of the two haloes has $V_c$=234 km/s and stellar and cold 
($T$~$\simeq$~10$^4$K) gas masses of
$M_{*}$=5.2$\cdot$10$^{10}$M$_{\odot}$ and 
$M_{\rm{cg}}$=0.9$\cdot$10$^{10}$M$_{\odot}$, respectively --- we shall 
denote this disk galaxy S1 in the following. The other disk galaxy formed
in the smaller of the two haloes has $V_c$=212 km/s, 
$M_{*}$=3.5$\cdot$10$^{10}$M$_{\odot}$ and 
$M_{\rm{cg}}$=0.6$\cdot$10$^{10}$M$_{\odot}$ at $z$=0 --- we denote it S2.
These and other physical parameters for the two galaxies are given in
Tables 1 and 2.

\begin{table*}
\begin{center}
\caption{Masses$^{\rm a}$ of and number of particles$^{\rm a}$ in selected disk
galaxies at $z$=0
\label{t:1}}
\small
\begin{tabular}{lcccccccc}
\hline\hline
Galaxy & $M_{*}$ & $M_{cg}$ $^{\rm b}$ & $M_{hg}$ $^{\rm c}$ &
 $M_{b,tot}$ $^{\rm d}$ & $N_*$ & $N_{cg}$ $^{\rm b}$ & $N_{hg}$ $^{\rm c}$ 
& $N_{b,tot}$ $^{\rm d}$\\
   & [10$^{10}$M$_{\odot}$] & [10$^{10}$M$_{\odot}$] & [10$^{10}$M$_{\odot}$]
& [10$^{10}$M$_{\odot}$] & & & &\\
\hline
S1 & ~~5.16 & 0.91 & 0.03 & 6.10 & ~~8493 & 1495 & ~54 & 10042\\
S2 & ~~3.46 & 0.61 & 0.03 & 4.10 & ~~5697 & 1001 & ~56 & ~~6754\\
S3 & ~~1.48 & 0.46 & 0.03 & 1.97 & ~~2431 & ~763 & ~47 & ~~3241\\
S3-8 & ~~1.65 & 0.20 & 0.03 & 1.88 & 21713& 2684 & 381 & 24778\\
\hline \hline
\multicolumn{9}{l}{$^{\rm a}$ Within galactocentric distance $r$=30 kpc}\\ 
\multicolumn{9}{l}{$^{\rm b}$ Gas with $T\le3\cdot 10^4$ K} \\
\multicolumn{9}{l}{$^{\rm c}$ Gas with $T>3\cdot 10^4$ K} \\
\multicolumn{9}{l}{$^{\rm d}$ Total baryonic mass and number of particles}
\end{tabular}
\end{center}
\end{table*}

\begin{table*}
\begin{center}
\caption{Other physical properties of selected disk galaxies at $z$=0
\label{t:2}}
\small
\begin{tabular}{lccccccccc}
\hline\hline
Galaxy & $V_c$ & $j_*$ & $j_{cg}$ & $B/D$ & $R_d$ & $\dot{M}_{gas}$ & SFR &
$b$ & Type\\
 & [km/s] & [kpc~km/s] & [kpc~km/s] & & [kpc] &
[M$_{\odot}$/yr] & [M$_{\odot}$/yr] & & \\
\hline
S1 & 234 & 532$\pm$21 & 1147 & 0.17$\pm$0.05 & 1.75$\pm$0.10 &
0.54 & 0.98 & 0.22 & Sab\\
S2 & 212 & 400$\pm$23 & ~387 & 0.27$\pm$0.08 & 1.43$\pm$0.07 &
0.71 & 1.70 & 0.62 & Sbc\\
S3 & 163 & 236 & ~664 & 0.28 & 1.28 & 0.30 & 0.22 & 0.19 & Sab\\
S3-8 & 166 & 248$\pm$20 & ~936 & 0.28$\pm$0.13 & 1.17$\pm$0.05
& 0.16 & 0.26 & 0.20 & Sab\\
\hline \hline
\end{tabular}
\end{center}
\end{table*}

In Figure~16 we show the rate at which cooled-out gas is accreted onto the
disk of S1 per radial bin at redshifts $z$=1 and $z$=0. The rate has been 
normalized
to unity over the entire disk at each redshift and is shown versus the radial 
coordinate in the
disk in units of the (stellar) disk scale length at $z$=0, $R_d$.
This disk is overall clearly forming ``inside-out'', but we note that already
at $z$=1 it is accreting gas out to $\sim$~6--8$R_d$, and at $z$=0 gas is 
still
accreting onto the inner parts of the disk. Figure~17 is similar to Fig.~16,
just for galaxy S2. It is apparent from the figure that the formation of this 
disk occurs in a strikingly different, ``outside-in'' manner. At
$z$=1 it is, like S1, already accreting gas out to $\sim$~6--8$R_d$, but at
$z$=0 the gas inflow has contracted to $R \la$~4$R_d$. Hence, though one
should obviously not draw any far reaching conclusions on the basis of just
two Milky Way like model disk galaxies, out results indicate that disk
formation (not surprisingly) is a more complicated process than depicted in 
the above mentioned analytical or semi-analytical works.

\begin{figure}
\plotone{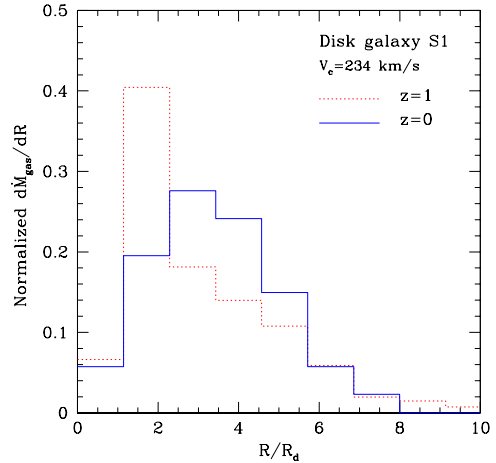}
\caption{Normalized disk gas accretion rates per radial bin at $z$=1 and $z$=0
for model disk galaxy S1 --- example of ``inside-out'' disk formation.
\label{f:16}}
\end{figure}

\begin{figure}
\epsscale{1.1}
\plotone{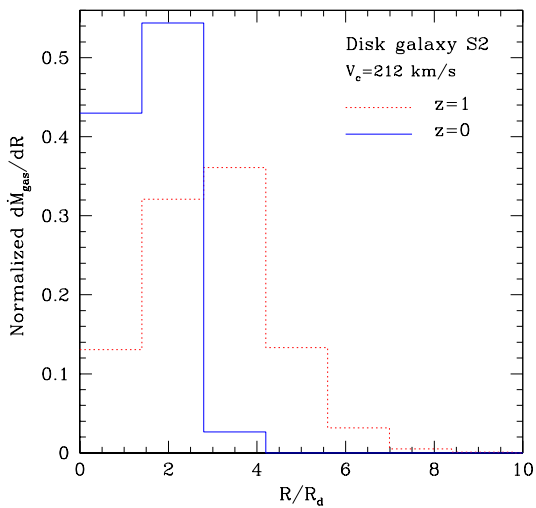}
\caption{Normalized disk gas accretion rates per radial bin at $z$=1 and $z$=0
for model disk galaxy S2 --- example of ``outside-in'' disk formation.
\label{f:17}}
\end{figure}

That gas is being deposited onto the disks out to 
{\mbox{$\sim$~6--8$R_d$}} is also
interesting in relation to the observational finding that present day HI disks
often extend well beyond the (stellar) Holmberg radius $R_{26.5}$
($\sim$~4--5~$R_d$ --- see van der Kruit \& Searle 1982), in fact
to about 1.5$R_{26.5}$ (e.g., Wolfe \etal 1986). We shall in a 
forthcoming paper discuss the implications of our models in relation to
observed medium to high-$z$ damped Ly$\alpha$ systems (DLAs).

The total gas accretion rates for galaxies S1 and S2 (integrated over the
entire disk) are 0.5 and 0.7 M$_{\odot}$/yr at $z$=0 and 3.0 and 4.5 
M$_{\odot}$/yr at $z$=1, so the gas accretion
rates are $\sim$~6--7 times larger at $z$=1 than at $z$=0. The masses of S1 and
S2 at $z$=1 are $M_{*}$=3.5 and 1.8 $\cdot$10$^{10}$M$_{\odot}$ and   
$M_{\rm{cg}}$=1.7 and 0.9$\cdot$10$^{10}$M$_{\odot}$, respectively. Given
the considerable drop in gas accretion rates from $z$=1 to $z$=0, that the 
galaxies
are accreting gas out to $\sim$~6--8$R_d$ at $z$=1 and the fairly modest
increase in total mass from $z$=1 to $z$=0, it seems safe to conclude that
the formation of isolated large disk galaxies was well under way by
redshift $z \sim$~1. This seems in line with the observations by Simard \etal
(1999), which indicate that disk galaxies have similar sizes at 
$z~\simeq$~1 and $z$=0. 

In Figure~18 we show for galaxies S1 and S2 the mean age of disk stars (within
2 kpc vertical distance from the disk) versus radial 
coordinate in units of the (stellar) disk scale length at $z$=0.
The bars correspond to the dispersion in each bin. Only stars
with formation redshift $z_f <$ 2.6 (corresponding to ages $\la$~12 Gyr)
were included --- this eliminates essentially all halo stars and also partly
the bulge stars for these two galaxies. In the disk of S1
there is essentially no age gradient and the mean stellar age is $\sim$~7--8
Gyr corresponding to a mean formation redshift of $\sim$~0.8 (in the two
innermost bins the mean age is larger due to contribution from
bulge stars). The disk
age dispersion is  $\sim$~2-3 Gyr corresponding to a dispersion in redshift
of $\sim$~0.3-0.4. At first it seems surprising that a disk, forming 
inside-out,
has no age gradient. However, firstly the star formation history is generally
different from the gas accretion history, and secondly the star formation
rate depends non-linearly on the cold gas density, $\dot{\rho}_{*} \propto
\rho_{\rm{cg}}^{1.5}$ (cf. eq.~2). This makes the star formation rate quite low
in the outer disk at late times and hence the average stellar age quite high
there (but some young stars are present --- see below).  

\begin{figure}
\epsscale{1.15}
\plotone{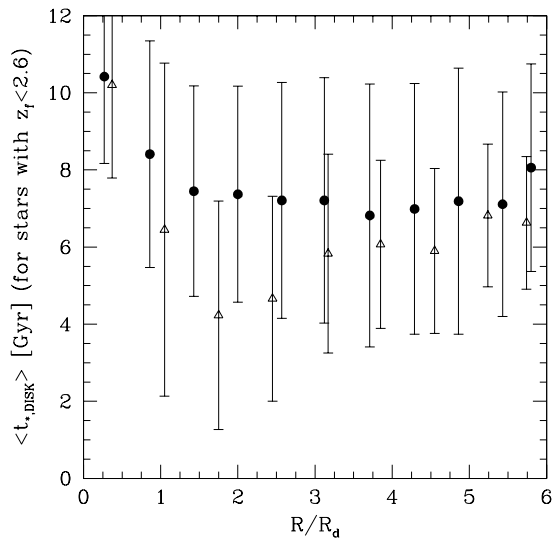}
\caption{Mean ages of disk stars in galaxies S1 (solid circles)
and S2 (open triangles). Only stars with formation redshift $z_f<$2.6
(or equivalently age $\la$~12 Gyr) are included. The bars show the age 
dispersions in each bin.
\label{f:18}}
\end{figure}

The mean age of the stars in the outer disk of galaxy S2 is $\sim$~6--7 Gyr
--- slightly less than the outer disk of S1. The mean age decreases with 
decreasing 
$R$ to about 4 Gyr in the third innermost bin. This is not 
surprising given 
the gas accretion patterns at $z$=1 and $z$=0 shown in Fig. 17. In the 
two innermost bins there is some
contribution from bulge stars (with $z_f <$~2.6) increasing the mean age there
to 6-10 Gyr. 

To gain more insight than what can be obtained from mean ages, we show in 
Figure~19 the distributions of stellar ages as a function of the radial 
coordinate in the two disks (again for stars with $z_f <$~2.6). The fractions 
of stars younger than 3, 6 and 9 Gyr are shown
for the disk of S1 with thick curves and for S2 with thin ones. 
For S1 the fractions
of stars younger than 3 and 6 Gyr increase steadily with $R$ out to 
$R~\sim$~4--5$R_d$. So the {\it distribution} of stellar ages does reflect that
the disk of S1 formed inside-out. At $R~\sim$~5--6$R_d$, the fractions of stars
younger than 3, 6 and 9 Gyr are $\sim$~15, 30 and 60\%.  

\begin{figure}
\epsscale{1.15}
\plotone{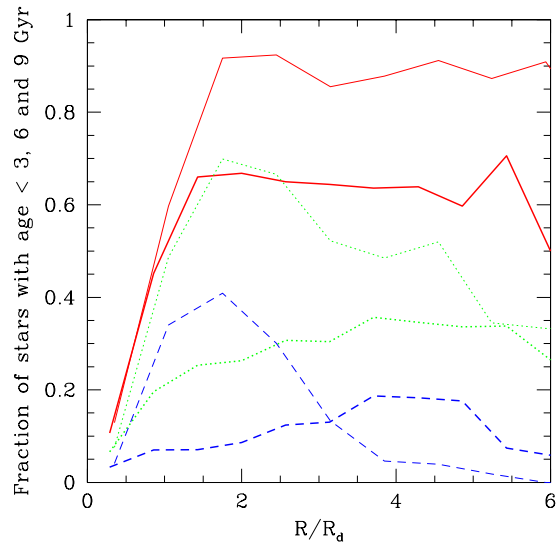}
\caption{Distribution of stellar ages in the disks of galaxies 
S1 (thick curves) and S2 (thin curves) as a function of the radial coordinate
in the disk. Only stars with $z_f<$2.6 (or equivalently age $\la$~12 Gyr) are
considered. The fraction of these younger than 3, 6 and 9 Gyr are 
marked by dashed, dotted and solid lines, respectively. The inner 
$\sim$~1.5$R_d$ are dominated by bulge stars.
\label{f:19}}
\end{figure}

The distribution of stellar ages in the disk of S2 reflects the outside-in
formation of this disk. At $R~\sim$~5--6$R_d$, there are almost no stars with
ages less than 3 Gyr, about 35\% with ages less than 6 Gyr, and only about
10\% with ages larger than 9 Gyr. The latter explains the slightly lower
mean age of the outer disk of S2 compared to S1.

Ferguson \& Johnson (2001) find in a recent observational study of the
outskirts of M31's disk that the typical stellar age at $\sim$~5--6 disk scale
lengths is $\ga$~7--8 Gyr. This seems in fair agreement with our results on the
mean stellar ages at $\sim$~5--6~$R_d$,
which are also lower limits since we have excluded all stars with $z_f >$~2.6.
But at least in the disk of S1 we find a non-negligible fraction of stars 
younger than 3 Gyr in the outskirts of the disk --- whether this is consistent
with the observational findings of Ferguson \& Johnson remains to be seen.

The mean age of the disk stars inside of $R=6R_d$, with $z_f<$~2.6 and 
excluding the two innermost bins, which have some bulge contribution, is
7.3 Gyr for S1. This corresponds to a mean
disk star formation redshift of $<\hspace{-1mm}z_f\hspace{-1mm}> \ga$~0.8, so 
for this galaxy even the
formation of the bulk of the {\it stellar} disk happens at $z \sim$~1. 
For galaxy S2,
the corresponding number is 4.9 Gyr, so for this galaxy the stellar
disk is arguably still forming --- its present SFR is also quite respectable, 
cf. Table 2. 

Finally, our results above on age gradients indicate that the colour gradients
observed in
galactic disks (with the disk generally becoming bluer with increasing $R$)
are a metallicity effect, rather than an age effect.

\subsection{Gas accretion histories for galactic disks and the ``G-dwarf
problem''}
Observational studies of the metallicity distribution of G and K stars in the
solar neighbourhood show that the distribution is very narrow (peaking around
{\mbox{[Fe/H]=--0.2~dex;}} Wyse \& Gilmore 1995; Rocha-Pinto \& Maciel 1996;
Hou \etal 1998; J{\o}rgensen 2000; Flynn \& Morell 1997; Kotoneva \etal 2002),
much too narrow to be consistent with what is expected from
a simple, closed-box model of galactic chemical evolution. This famous, 
apparent inconsistency has been coined the ``G-dwarf problem'' and was first
discussed by van den Bergh (1962), Schmidt (1963) and Pagel \& Patchett (1975).
The currently most popular (and very sensible --- 
see below) answer to the G-dwarf problem is to assume that the solar 
neighbourhood was formed gradually over the Hubble
time through prolonged infall of fairly low metallicity cooled-out gas
originating from the hot halo; such ``infall models'' were first introduced
by Larson (1972) and Lynden-Bell (1975). Assuming, as is often 
done
for simplicity, that the infall rate is an exponentially decaying
function of time an infall timescale of $t_{inf}$=7--10~Gyr is required to get
the models to match recent observations (Chiappini \etal 1997; Portinari \etal
1998; Boissier \& Prantzos 1999). If a gaussian form for 
the infall rate is assumed, then the typical timescale is $\sim$5~Gyr 
(Prantzos \& Silk 1998; Chang \etal 1999).
With the current models we are able to test these assumptions: 

In Figure~ 20
and 21 we show the histories of accretion of cooled-out gas for the disks of
galaxies S1 and S2, respectively. In each figure we show by solid line the 
accretion rate history of the
entire disk and by dotted that of the ``solar cylinder''
(operationally defined as 2.1$R_d$$\le$$R$$\le$3.2$R_d$; we assume that the 
scalelength of the Galactic disk is 3.0 kpc, as a compromise between various
recent determinations, and that the solar circle is at $R_0$=8.0 kpc).
For S1 the rate of gas infall onto the ``solar cylinder'' is almost perfectly
exponentially decaying with an e-folding time of $\sim$~5 Gyr (the
earliest bin in the figures correspond to $z$=2.2). For S2 the infall history
looks more like the one derived by Prantzos \& Silk (2003) for the
solar cylinder using a combination of observational data and theoretical
arguments. From a linear fit to the infall rate history of the ``solar 
cylinder'' over the age range 0--12 Gyr
we derive an infall timescale of $\sim$~6 Gyr. So the infall timescales
derived fall short of the 7--10 Gyr required by current chemical evolution 
models, but only marginally so. Moreover, we remind the
reader again that we have been analyzing just two Milky Way like
disk galaxies in detail.

\begin{figure}
\epsscale{1.04}
\plotone{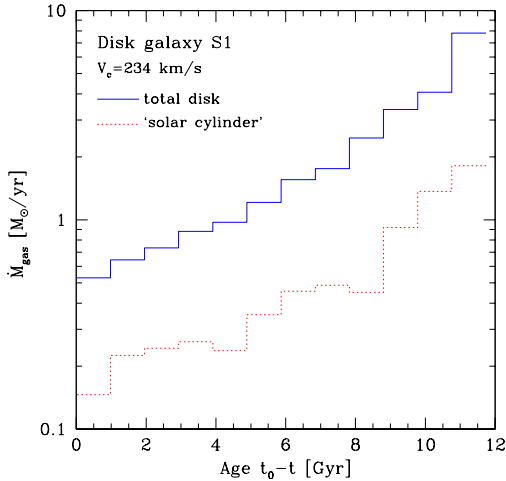}
\caption{Rates of infall of cooled-out halo gas onto the disk
of galaxy S1 versus age. Solid line: all disk, dotted line: ``solar cylinder''
(operationally defined as 2.1$R_d$$\le$$R$$\le$3.2$R_d$).
\label{f:20}}
\end{figure}

\begin{figure}
\epsscale{1.04}
\plotone{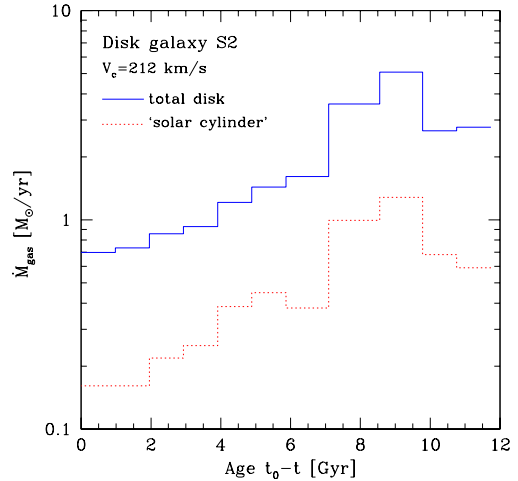}
\caption{As Fig. 20, but for galaxy S2.
\label{f:21}}
\end{figure}

\subsection{Hot gas in galaxy haloes and pulsar dispersion measures}
The amount of hot gas ($T \sim 10^6$ K) in the Galactic halo can be 
constrained from 
dispersion measures to pulsars in the Galactic halo. Of particular
interest is one pulsar in the high latitude globular cluster M53 and three
in the LMC (Large Magellanic Cloud), because the distances to these pulsars
are known. Rasmussen (2000) finds from these data and some modelling
of the hot halo an upper limit on the cumulative amount of hot gas to the
distance of the LMC of 1.5--2$\cdot$10$^9$ M$_{\odot}$. In Figure~22 we show
the cumulative mass of hot gas vs. galactocentric distance
for three Milky Way like disk galaxies. Solid curves are for gas with 
$T>10^5$ K, dotted for {\mbox{$T>10^6$ K.}}
Top set of curves are for disk galaxy S2, middle for
S1 and lower are for a simulation of SPSF type of the halo corresponding to
galaxy S1, but with a gas abundance of 1/3 solar (see Toft \etal 2002),
rather than the primordial gas abundance used in this work. It is clear that
all three hot gas haloes satisfy the above constraint increasingly so with
increasing gas abundance, as discussed by Toft \etal 

\begin{figure}
\epsscale{1.04}
\plotone{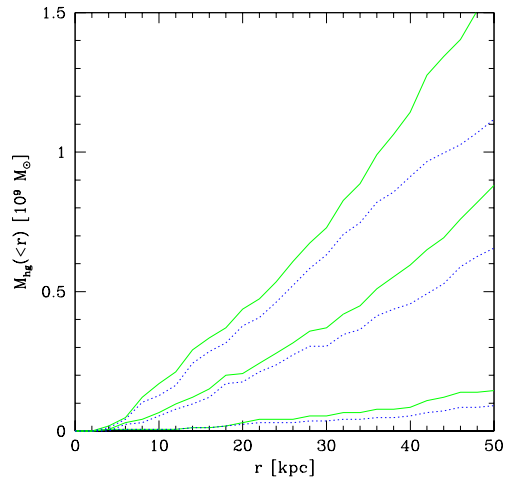}
\caption{Cumulative mass of hot gas vs. galactocentric distance for three 
Milky Way like galaxies. Solid curves are for gas with {\mbox{$T>10^5$ K,}}
dotted for 
$T>10^6$ K. Top set of curves are for disk galaxy S2, middle for S1 and lower 
are for a simulation of SPSF type of the halo corresponding to galaxy S1, but 
with a gas abundance of 1/3 solar (see Toft \etal 2002), rather than 
the primordial gas abundance used in this work. 
\label{f:22}}
\end{figure}

With the present models we can, however, do better than this, because we can
``insert'' M53 and the LMC in our model galaxy haloes and calculate the
expected dispersion measures directly:

For electromagnetic waves propagating through a plasma, the group velocity
$v_{gr}$ will depend on the frequency $\nu$ of the radiation. Specifically,
for radio waves in an unmagnetized plasma (e.g., F\"althammer 1995)
\begin{equation}
v_{gr} = c~\sqrt{(1-\frac{\nu_{pe}^2}{\nu^2})} ~~,
\end{equation}
where $\nu_{pe}=(n_e e^2/\pi m_e)^{1/2}$ is the electron plasma
frequency, $e$ the electron charge, $m_e$ the mass of the 
electron and $c$ the speed of light. A pulse travelling a distance $L$ in the 
plasma will have a travel time $t$ given approximately by
\begin{equation}
t \approx \frac{L}{c} + \frac{e^2 D_m}{2 \pi m_e c \nu^2}
\end{equation}
rather than the ``vacuum value'' $L/c$ (e.g., Spitzer 1978), where the
{\it dispersion measure} $D_m$ is defined as 
\begin{equation}
D_m \equiv \int_0^L n_e dl ~~.
\end{equation}
Detected radio signals from pulsars show a frequency-dependent delay that can
be used to evaluate $D_m$. For the pulsar in the globular cluster M53, located
at a distance of $\approx$18.9 kpc at $(l,b)$ = (333$^{\circ}$,80$^{\circ}$) 
one finds $D_m$=24.0 
cm$^{-3}$~pc. For the three pulsars in the LMC, located at a distance of 
$\approx$49.4 kpc at $(l,b) \simeq$ (279$^{\circ}$,-33$^{\circ}$), 
$D_m$=115$\pm$13 cm$^{-3}$~pc is found (Rasmussen 2000).

To calculate what values of $D_m$ we predict to M53 and the LMC for the 
haloes of S1 and S2 we average over the hot halo gas free electron densities
in spherical shells and correct the densities for  
resolution effects at the disk-halo interface. We calculate $D_m$ as the
average of the value at $z$=0 and the values obtained at times 100 and
200 Myr prior to this. One has to adopt a value for $R_0$, the distance of
the sun from the galactic center, when calculating $D_m$ for the model 
galaxies. Though the {\it stellar} disks are too small compared to reality, 
this is not necessarily the case for the hot halo. We hence did the 
calculations of $D_m$ with two values of $R_0$=5 and 8 kpc. For the M53
pulsar we find for S1 24$\pm$2 and 16$\pm$1 and for S2 28$\pm$1 
and 17$\pm$1 cm$^{-3}$~pc for $R_0$=5 and 8 kpc, respectively, in fairly good 
agreement with the observed value. For the LMC we find
for S1 26$\pm$2 and 18$\pm$1 and for S2 31$\pm$1 
and 20$\pm$1 cm$^{-3}$~pc for $R_0$=5 and 8 kpc, respectively. These values are
considerably smaller than the observed $D_m$ --- this is probably due to 
additional contributions to the observed $D_m$ from 
the Gum Nebula, a largely ionized region at $l \sim$ 
230$^{\circ}$-290$^{\circ}$ and distance of 
about 0.5 kpc (Taylor \& Cordes 1993), and the diffuse, ionized 
Reynolds-Layer (Reynolds 1993) in the direction towards the LMC
(Rasmussen 2000).

It is clear from Fig.~22 and the discussion in Toft \etal (2002) that
the predicted values of $D_m$ above are upper limits in the sense that had we
included chemical evolution in the simulations it would have lead to some
non-zero level of metal enrichment in the hot halo gas, which in turn would
have caused a reduction in the predicted values of $D_m$.  

\subsection{Star formation histories}
As an example of the typical star formation histories in our simulations we
show (for the {\it same} dark matter halo) in Figure~23 the star formation 
rate in units of the current one for
three of the five simulations resulting in the largest disk galaxy (with
$V_c \simeq$ 220--230 km/s). Two are with self-propagating star formation
with $n_{H,e,low}$=0.05 and 0.25 cm$^{-3}$ and one is for the simulation
without self-propagating star formation. All three reproduce the observed
peak in the SFR history at $z$$\sim$2 (Madau \etal 1996), but whereas 
the simulations with
self-propagating star formation have a low at $z$$\sim$4.5--5.5 and another
peak at $z$$\sim$6--8 corresponding to the putative population III, the one 
without self propagating star formation has
a monotonically decreasing SFR with increasing redshift. 
Though it would obviously be premature to take the star formation 
histories of our simulations of the formation and evolution of {\it individual}
and fairly large galaxies as representative of the SFH of the Universe as a 
whole, our results suggest that the global SFH could possibly be bimodal.
In relation to this it is interesting that recent observations of high-$z$ 
quasars may indicate that the Universe was reionized at a 
redshift of about 6 (Becker \etal 2001; Djorgovski \etal 2001;
but see also Hu \etal 2002, who advocate $z_{reion}\ga$~6.6 from 
observations of a candidate $z$=6.56 galaxy; this is, however, still consistent
with the location of the second SFR peak discussed above).
The different SFH
scenarios should be observationally constrainable with upcoming instruments
like JWST and ALMA. 

\begin{figure}
\plotone{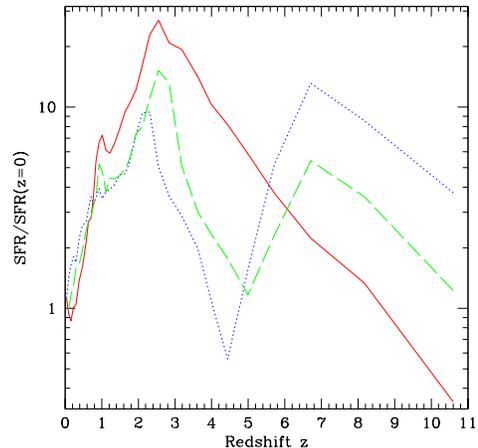}
\caption{Star formation histories in three of the five of our simulations 
forming the largest disk galaxy (with $V_c \simeq$ 220-230 km/s). Two are 
for self-propagating star formation simulations
with $n_{H,e,low}$=0.05 and 0.25 cm$^{-3}$, and are shown by dotted and 
dashed lines respectively. The last one is for the simulation
without self-propagating star formation and is shown by the solid line.
\label{f:23}}
\end{figure}

\subsection{Effects of numerical resolution}
To check for effects of numerical resolution we did the following two 
tests: 

1. We ran a series of 7 SPSF disk galaxy formation simulations with 
$n_{H,e,low}$=0.2 and with gravitational softening lengths equal to half the 
``standard'' values, i.e. 
{\mbox{0.65 $h^{-1}$kpc}} and 1.4 $h^{-1}$kpc for the SPH (and star) and
active DM particles, respectively. The median normalized stellar specific
angular momentum was 11$\pm$15\% lower than for the ``standard'' simulations.
This was mostly due to the median $V_c$ being 4$\pm$12\% larger than for the
standard simulations. So there may be a small (and statistically 
insignificant) effect of the gravitational softening on the resulting
normalized specific angular momenta - but see below. 

2. The second test was to re-simulate one of our disk galaxies, which we
shall denote S3 in the following, at 8 times higher mass and 4 times higher 
force resolution, as detailed at the end of section 4. We shall denote the 
higher mass and force resolution version of S3 by S3-8 in the following. 
Masses, numbers of particles and other physical characteristics of S3 and S3-8 
at $z$=0 are given in Tables 1 and 2. The total (baryonic) masses of S3 and 
S3-8 are
very similar, differing by less than 5\%. The stellar mass of S3-8 is 11\%
larger than that of S3 and the cold gas mass correspondingly smaller. We
believe that it is the combination of the ability to resolve gas clouds of
smaller masses and the additional small scale power that leads to the somewhat
higher stellar mass and correspondingly lower cold gas mass of the high
resolution galaxy S3-8 relative to S3. The characteristic circular speed of
S3-8 is just 2\% larger than that of S3 despite the 4 times higher force
resolution. This may indicate that force resolution used in our ``standard''
simulations is adequate. The bulge-to-disk ratio of S3-8 is B/D=0.28$\pm$0.13.
To obtain the $j_*$ and $R_d$ given in Table 2 for S3 we forced the 
bulge-to-disk decomposition to have B/D=0.28 for S3 to enable a fair 
comparison between S3 and S3-8. We find that $j_*$ for S3-8 is 5$\pm$8\% 
larger than that of S3 and that the specific angular momentum of the cold gas
$j_{cg}$ is 41\% larger. The gas accretion rate is somewhat lower (this is
presumably related to the larger stellar mass and smaller cold gas mass of
S3-8 relative to S3) and the star formation rates and birthrate parameters, 
$b$, are very similar.

We conclude that the high mass resolution simulation generally is in good 
agreement
with the ``standard'' resolution one, with a tendency for the stellar mass to
be somewhat larger, the cold gas mass somewhat smaller and the specific angular
momenta somewhat larger. All in all it is hence possible that increased 
mass resolution will be beneficial also with respect to the AM problem.
\setcounter{footnote}{0}
\section{Conclusion and outlook}
\label{s:conclusions}
In conclusion we have obtained a mix of realistic disk, lenticular and 
elliptical galaxies\footnote{Pictures of some of the galaxies can be seen at 
http://www.tac.dk/\~~\hspace{-1.4mm}jslarsen/Hubble\hspace{0.1mm}\_ \hspace{-0.7mm}Sequence}
in our $\Lambda$CDM galaxy formation simulations with effects of
energetic stellar feedback processes included.
We find that the disk galaxy angular momentum problem can be considerably 
alleviated (though not entirely solved as yet) in this way, provided that the
early star-bursts are fairly strong, converting 2-5\% of the initial gas mass
into stars. For early star-burst strengths in this range, the improvement on 
the AM problem is about the same, so extreme fine-tuning of the feedback seems
not required.

The stellar disks have approximately exponential surface density profiles and
those of the bulges range from exponential to $r^{1/4}$, as observed. 
The bulge-to-disk ratios of the disk galaxies are consistent
with observations and likewise are their integrated $B$--$V$ colours, which 
have
been calculated in a simplified way (``post-process'') using stellar 
population synthesis techniques. 
Furthermore we can match the observed $I$-band Tully-Fisher 
relation, provided that the stellar mass-to-light ratio of disk galaxies is
$M/L_I \sim$ 0.8, similar to what was found by Sommer-Larsen \& 
Dolgov (2001) from their WDM simulations and in fair agreement with 
several recent observational determinations of $M/L_I$ for disk galaxies.

The ellipticals and lenticulars have approximately $r^{1/4}$ stellar surface
density profiles, are dominated by non disk-like kinematics and flattened due 
to non-isotropic, stellar velocity distributions, again consistent with 
observations.

We predict that hot halo gas is cooling out and being accreted onto the 
Galactic disk at a rate of {\mbox{0.5--1 M$_{\odot}$/yr}} at $z$=0, consistent
with upper limits deduced from $FUSE$ observations of \osix. 
We have analyzed the formation history of two Milky Way like disk galaxies in
detail and find gas accretion rates, and
hence bolometric X-ray luminosities of the haloes, 6--7 times
larger at {\mbox{$z \sim$1}} than at $z$=0 for these disk galaxies. More 
generally, we find that gas infall rates onto these disks are nearly
exponentially declining with time, both for the total disk and the ``solar 
cylinder''.
This theoretical result hence supports the exponentially declining gas infall
approximation often used in chemical evolution models. The infall time-scales 
deduced are $\sim$5--6 Gyr, comparable to what
is adopted in current chemical evolution models to solve the ``G-dwarf
problem''.

It is moreover found for these two galaxies that galactic disks can form 
``inside-out'' as well as
``outside-in'', but in both cases the mean ages of the stars in the outskirts
of the disks are $\ga$6--8 Gyr, broadly consistent with the findings of
Ferguson \& Johnson for the disk of M31.

The amount of hot gas in disk galaxy haloes is consistent with observational 
upper limits. We ``insert'' the globular cluster M53 and the LMC in the haloes
of the two Milky Way like disk galaxies and calculate dispersion measures to 
these
objects. Our results are consistent with upper limits from observed 
dispersion measures to pulsars in these systems. 

Our simulations indicate that we can reproduce the observed peak
in the cosmic star formation rate at redshift $z \sim 2$.
Depending on the star formation and feedback scenarios
we predict either a monotonically decreasing cosmic star formation rate beyond
these redshifts or a second peak at $z$$\sim$6--8 corresponding to the
putative population III and interestingly similar to recent estimates of the
redshift at which the Universe was reionized. These various scenarios
should hence be observationally constrainable with upcoming instruments
like JWST and ALMA.    

In this paper we have mainly been concerned with the present day properties
of the simulated disk galaxies. We shall in forthcoming papers study the 
present day properties of the E/S0s in detail as well as the dynamics of all
types of galaxies. In particular, we intend to study the kinematics and 
dynamics of the stellar haloes of Milky Way like disk galaxies to compare
with current models of the Galactic stellar halo (e.g., Sommer-Larsen \etal
1997; Helmi \& White 1999; Helmi \etal 1999). We will 
also study the higher
redshift properties of all types of galaxies and the properties of medium to 
{\mbox{high-$z$}} damped Ly$\alpha$ systems predicted by our simulations. 
Moreover, we will incorporate chemical evolution including
non-instantaneous gas and heavy element recycling in the simulations and
replace thermal energy by entropy as an independent variable in SPH (Springel
\& Hernquist 2002).

Last, but not least we will also study
much more closely the physical factors determining the final Hubble type of
the galaxies formed. It is, however, already at this point clear from our
simulations that
final galaxy morphologies primarily reflect the merging histories of the 
galaxies in the sense of the very {\it detailed} merging histories --- just to
give an example, we
find that even major merging sometimes can be {\it constructive} in forming 
disk galaxies with extended disks. Another interesting result is that whereas
the mean spin-parameter for the 5 spheroid dominated systems (E/S0) is
$<\hspace{-1mm}\lambda_{200}\hspace{-1mm}>_{E/S0}$=0.043$\pm$0.005, it is 
$<\hspace{-1mm}\lambda_{200}\hspace{-1mm}>_S$=0.037$\pm$0.002 for the 35 disk 
galaxies. Hence our
results do not seem to support the notion that the DM haloes in which disk 
galaxies form on average have larger spin-parameters than the haloes of early
type galaxies (e.g., Mo, Mao \& White 1998;
van den Bosch 1998).\\[.5cm] 

We have benefited from discussions with G.~Bryan, Y.-I.~Byun, G.~Carraro,
C.~Chiappini, C.~Chiosi, P.~R.~Christensen,
R.~Dominguez-Tenreiro, A.~Ferguson, C.~Flynn, K.~Freeman, J.~Hjorth,
R.~Kennicutt, A.~Kravtsov, C.~Lia, F.~Matteucci, J.~Monaghan, 
B.~Moore, {\AA.}~Nordlund, B.~Pagel, K.~Pedersen, N.~Prantzos, V.~Quilis, 
J.~Rasmussen,
A.~Saiz, B.~Savage, V.~Springel, S.~Toft, G.~Yepes, S.~Yi and, in particular, 
Joe Silk. The comments by the anonymous referee significantly improved the
presentation of our results. LP acknowledges
kind hospitality from the Astronomy Department of Padua and from the 
Observatory of Helsinki. JSL acknowledges the help by A.~Sommer-Larsen and
M.~Vesterg{\aa}rd in analyzing the observational data used in this paper.
This work was supported by Danmarks Grundforskningsfond through its support
for the establishment of the Theoretical Astrophysics Center.

\end{document}